\documentclass[lettersize,journal]{IEEEtran}
\usepackage{amsmath,amsfonts}
\usepackage{mathrsfs}
\usepackage{algorithmic}
\usepackage{algorithm}
\usepackage{array}
\usepackage[caption=false,font=normalsize,labelfont=sf,textfont=sf]{subfig}
\usepackage{textcomp}
\usepackage{stfloats}
\usepackage{url}
\usepackage{verbatim}
\usepackage{xcolor}
\usepackage{graphicx}
\usepackage{cite}
\usepackage{booktabs}
\usepackage{makecell}
\usepackage{multirow}
\usepackage{graphicx}
\begin{document}

\title{Physics-Informed Spatial-Temporal Transformer for Terahertz Near-Field Beam Tracking}

\author{Zhi Zeng,~\IEEEmembership{Student Member,~IEEE}, Chong Han,~\IEEEmembership{Senior Member,~IEEE}, and Emil Björnson,~\IEEEmembership{Fellow,~IEEE}
\thanks{An earlier version of this paper will be presented in part at the IEEE ICASSP, May 2026~\cite{zeng2026}.}
\thanks{Zhi Zeng is with the Terahertz Wireless Communications (TWC) Laboratory, Shanghai Jiao Tong University, Shanghai 200240, China (e-mail: zhi.zeng@sjtu.edu.cn).}
\thanks{Chong Han is with the Terahertz Wireless Communications (TWC) Laboratory and also the Cooperative Medianet Innovation Center (CMIC), School of Information Science and Electronic Engineering, Shanghai Jiao Tong University, Shanghai 200240, China (e-mail: chong.han@sjtu.edu.cn).}
\thanks{Emil Björnson is with the Department of Communication Systems, KTH Royal Institute of Technology, 100 44 Stockholm, Sweden (e-mail: emilbjo@kth.se).}}

\maketitle

\begin{abstract}

Terahertz (THz) ultra-massive multiple-input multiple-output (UM-MIMO) promises ultra-high throughput, while its highly directional beams demand rapid and accurate beam tracking driven by precise user-state estimation.
Moreover, large array apertures at high frequencies induce near-field propagation effects, where far-field modeling becomes inaccurate and near-field parametric channel estimation is costly.
Bypassing near-field codebook, PAST-TT is proposed to bridge near-field tracking with low-overhead far-field codebook probing by exploiting parallax, amplified by widely spaced subarrays. With comb-type frequency-division multiplexing pilots, each subarray yields frequency-affine phase signatures whose frequency and temporal increments encode propagation delay and its variation between frames. 
Building on these signatures, a Parallax-Aware Spatial Transformer (PAST) compresses them and outputs per-frame position estimates with token reliability to downweight bad frames, regularized by a physics-in-the-loop consistency loss. A causal Temporal Transformer (TT) then performs reliability-aware filtering and prediction over a sliding window to initialize the beam of the next frame. Acting on short token sequences, PAST-TT avoids a monolithic spatial-temporal network over raw pilots, which keeps the model lightweight with a critical path latency of 0.61 ms. Simulations show that at 15~dB signal-to-noise ratio, PAST achieves 7.81~mm distance RMSE and 0.0588$^\circ$ angle RMSE. Even with a bad-frame rate of 0.1, TT reduces the distance and angle prediction RMSE by 23.1\textbf{\%} and 32.8\textbf{\%} compared with the best competing tracker.

\end{abstract}


\section{Introduction}
\label{sec:I}

\IEEEPARstart{T}{erahertz} (THz) communications have emerged as an attractive component for future wireless systems, as the ultra-broad spectrum from 0.1~THz to 10~THz provides abundant bandwidth and opens the door to extreme-capacity wireless links~\cite{shafie2022terahertz}. Nevertheless, the notable path loss in the THz band can affect the link budget and limit coverage~\cite{akyildiz2018combating}. Fortunately, the short wavelength enables dense integration and ultra-large apertures, making ultra-massive multiple-input multiple-output (UM-MIMO) a natural companion to THz for harvesting substantial beamforming gains and compensating for the severe propagation loss~\cite{han2018ultra}. To make THz UM-MIMO practical, hybrid beamforming (HBF) with a limited number of RF chains is typically adopted for hardware efficiency, which constrains how many beams can be probed simultaneously and makes low-overhead training imperative~\cite{yan2020dynamic,yan2021joint}. However, the resulting pencil beams are highly sensitive to mobility and channel dynamics, and even modest pointing errors can cause severe throughput degradation~\cite{chen2021millidegree,chen2024can,11239413}. Consequently, accurate beam alignment and tracking become essential, requiring frequent beam updates based on the time-varying user state under tight per-frame latency and training overhead constraints.

Furthermore, this alignment requirement becomes more challenging because large array apertures at high carrier frequencies confine THz links into the radiative near-field (NF), where the conventional far-field (FF) plane-wave model (PWM) is inaccurate due to non-negligible wavefront curvature, whereas element-wise spherical-wave model (SWM) can be computationally prohibitive for UM-MIMO~\cite{chen2024can,11239413,cui2022channel,lu2023hierarchical,chen2023beam}. Additionally, in THz links with a dominant line-of-sight (LoS) path, NF beams depend jointly on angle and distance, which substantially enlarges the search space for beam management and makes joint angle-distance sweeping prohibitively expensive for low-latency tracking~\cite{cui2022near,cui2022channel}. Meanwhile, practical training observations can be unreliable due to intermittent blockage and other impairments, further destabilizing online tracking if not properly detected and handled.

Taken together, these considerations underscore an urgent need for NF beam tracking mechanisms that simultaneously achieve high accuracy, strong robustness to occasional unreliable frames, and low control latency under practical THz UM-MIMO architectures.

\subsection{Related Work}
\label{ssec:IA}

Extremely large apertures at THz bands move a non-negligible portion of propagation from purely FF into the radiative NF, where wavefront curvature breaks the FF plane-wave abstraction and makes beamforming and channel acquisition jointly depend on angle and distance~\cite{chen2024can,11239413,cui2022channel,lu2023hierarchical,chen2023beam}.
Accordingly, extensive efforts revisit beam training and alignment under NF channel models and polar-domain representations that parameterize paths by direction and distance~\cite{chen2024can,11239413,cui2022channel,lu2023hierarchical,chen2023beam,chen2024triple,you2023near}.
To reduce the prohibitive two-dimensional search burden, hierarchical training and structured NF codebooks have emerged, including spatial-chirp~\cite{shi2023spatial} and multi-stage refinement strategies that progressively narrow down the candidate region in the angle-distance domain~\cite{chen2024triple,lu2023hierarchical}.
Related works also consider staged refinement that first leverages FF-like probing to confine angular candidates and then applies NF-aware processing to refine distance parameters, aiming to balance modeling accuracy and training complexity~\cite{zhang2022fast}.

In addition, polar-domain codebook size and search overhead can scale sharply with the number of antennas, mobility and HBF constraints, which becomes particularly challenging for per-frame beam updates under tight control-latency budgets~\cite{you2023near}.
Beyond one-shot alignment, NF beam tracking has also started to attract attention, for example, by combining kinematic modeling with recursive estimators under hybrid architectures~\cite{chen2023beam}.
Nevertheless, many NF beam-management approaches still rely on searching over a large polar-domain candidate set to resolve distance-dependent focusing~\cite{you2023near,lu2023hierarchical}.  
This leaves a gap for fast beam management: \textit{extracting NF geometric evidence under a lightweight probing budget without committing to NF codebook sweeping}. 

In practical THz UM-MIMO deployments, unified probing over a wide distance range is favored, motivating NF tracking designs to operate with FF discrete Fourier transform (DFT) codebook probing~\cite{giordani2018tutorial}.
Specifically, several works revisit the long-standing assumption that FF codebooks are unsuitable in the NF, and show that FF DFT probing can still expose distance-sensitive signatures through beam-response patterns, enabling NF alignment with reduced overhead~\cite{chen2024can,11239413,JARE}.
In parallel, widely-spaced multi-subarray (WSMS) architectures have been advocated as practical THz front-ends that preserve high per-subarray gain while exploiting inter-subarray phase diversity, thereby enriching NF geometric information compared to co-located arrays~\cite{yan2021joint}.
Performance analyses further quantify such benefits by relating angle-distance accuracy to the multi-view aperture created by the subarray layout~\cite{yang2024performance}.
Closely related, position- or geometry-aided beam management leverages user-state information to reduce beam selection overhead in highly directional systems~\cite{garcia2018transmitter}.

Despite these advances, existing solutions are often developed for one-shot alignment or isolated estimation tasks, and it remains underexplored how to convert low-overhead probing outcomes into a persistent and trackable measurement representation that can be robustly consumed by an online tracker under intermittent impairments.
This motivates a second gap: \textit{a structured handoff from FF probing to NF tracking that carries forward geometry evidence together with a principled notion of measurement reliability}.

After initial access, beam tracking aims to maintain alignment for mobile users with minimal overhead by exploiting temporal correlation in positions, path gains, or other low-dimensional channel parameters~\cite{yi2024beam}. Kalman-type trackers perform recursive prediction and correction using sounding-beam observations, and robust extensions add realignment triggers and outlier rejection to mitigate loss of track under blockage and intermittent observations~\cite{EKF, larew2019adaptive, zhang2016tracking, wang2018robust}.
In parallel, deep learning (DL) has been increasingly explored for beam tracking, where sequence models map historical pilots or channel state information to future beams~\cite{LSTM}. Beyond end-to-end predictors, learning-augmented filtering preserves the recursive Bayesian prediction-update structure, while using neural networks to learn key components that are difficult to model accurately, such as the Kalman gain~\cite{kalmannet}. More recently, DL has also been applied to NF and UM-MIMO beam training and tracking, typically by CNN-assisted hierarchical training and RNN-based tracking coupled with sweeping~\cite{wang2025near,dehkordi2021adaptive}. Nevertheless, many learning-based designs still treat measurements as high-dimensional black box inputs or rely on additional probing to explore the joint angle-distance space.

While these methods can improve accuracy in challenging dynamics, online deployment in highly directional THz systems still relies on strict causality, robustness against sporadic bad frames and tight inference latency within millisecond-scale control loops. Moreover, much of the tracking literature is rooted in FF angle-only abstractions~\cite{zhang2016tracking,larew2019adaptive}, whereas NF tracking involves angle-distance coupling and regime-dependent measurement structures. Therefore, it remains underexplored to address the third gap: \textit{an end-to-end NF tracking loop that is explicitly reliability-aware and latency-conscious under practical THz beam management budgets}.

\vspace{-0.1\baselineskip}
\subsection{Contributions}
\label{ssec:IB}


In this work, we enable THz NF beam tracking through parallax under lightweight FF DFT probing.
To form a tracking-ready interface from comb-type frequency-division multiplexing (FDM) pilots, we design a parallax-aware spatial transformer (PAST) for one-shot NF localization and reliability-aware tokenization. Then, we develop a causal temporal transformer (TT) to close the online beam tracking loop, leading to PAST-TT.
The main contributions are summarized as follows.

\begin{itemize}
  \item \textbf{We establish a parallax bridge that enables NF tracking under a lightweight FF probing budget.}
  Under a widely-spaced-subarray architecture with comb-type FDM pilots, we consider a hybrid spherical- and planar-wave channel model (HSPM).
  The phase signature of each subarray exhibits an approximately frequency-affine structure, and its adjacent-tone and inter-frame increments encode the propagation delay and its variation between frames.
  This leads to structured space-frequency-time evidence that supports low-overhead geometric inference without exhaustive joint angle-distance sweeping.

  \item \textbf{We propose the PAST as a physics-informed measurement encoder with explicit reliability-aware tokenization.}
  Leveraging the above evidence structure, high-dimensional pilots are distilled into compact physical tokens that preserve parallax and provide reliability statistics, suppressing intermittently corrupted observations without relying on fragile phase unwrapping.
  Geometry-biased, reliability-gated attention and a physics-in-the-loop consistency regularizer jointly improve the localization accuracy and measurement quality for each frame.

  \item \textbf{We develop the TT 
  as a causal reliability-aware filter-predictor to close the tracking loop under bad frames.}
  Using the per-frame tokens from PAST, TT performs reliability-injected masked attention over a sliding window and outputs a filtered state and a one-step prediction, where the prediction directly initializes the next-frame probing beam under the same FF DFT codebook.
  An
  increment descriptor and a temporal physics loop further mitigate error propagation under bad frames.

  
  \item \textbf{We evaluate the performance of PAST-TT compared with other representative methods.}
  We carry out extensive simulations across SNRs, bad-frame rates and mobility. The results demonstrate that PAST-TT outperforms the compared solutions with improved estimation accuracy and robustness under intermittently corrupted observations, leading to higher spectral efficiency. Complexity analysis and GPU latency measurement further confirm that its critical-path latency at each frame stays below the frame interval and enables real-time tracking.

\end{itemize}

The remainder of this paper is organized as follows. Sec.~\ref{sec:II} presents the system model with the parallax-aware NF channel and signal structure, and the problem formulation is also presented. Sec.~\ref{sec:III} introduces PAST for per-frame localization and reliability-aware evidence extraction (an earlier version of the NF localization part was presented in~\cite{zeng2026}). Sec.~\ref{sec:IV} develops the causal TT for online filtering and one-step prediction. Sec.~\ref{sec:V} evaluates the performance of the proposed methods. Finally, Sec.~\ref{sec:VI} concludes the paper.

\section{System Overview and Problem Formulation}
\label{sec:II}
In this section, we first introduce the system model of THz UM-MIMO, followed by the HSPM with parallax effects and the observation signal structure. Then, we formulate the per-frame localization and online beam tracking problems.

\begin{figure}[t]
\centering
\includegraphics[width=0.476\textwidth,trim=0cm 0cm 0.05cm 0cm,clip]{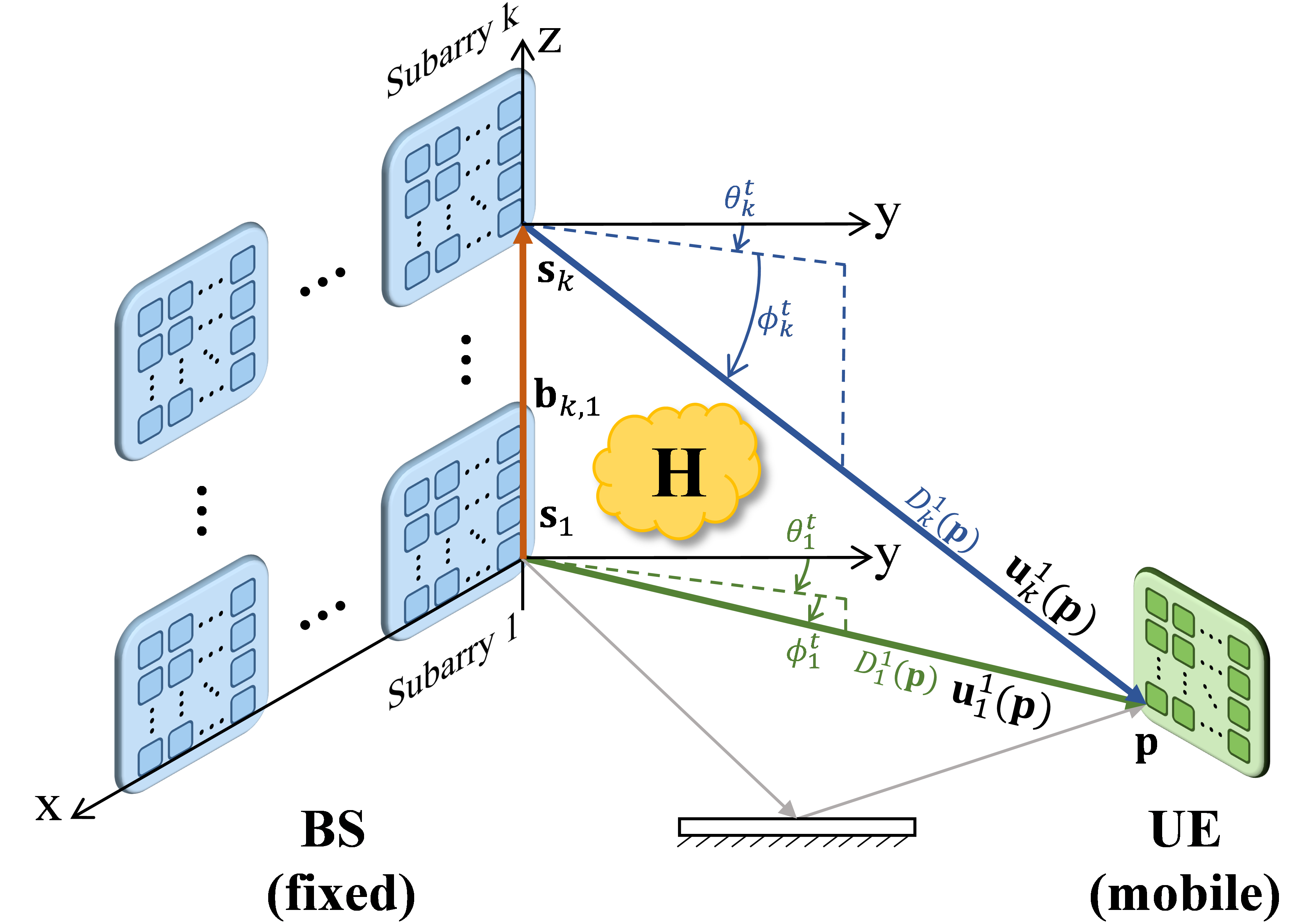}
\caption{THz UM-MIMO system model with fixed BS and mobile UE.}
\label{fig1}
\end{figure}

\vspace{-0.3\baselineskip}
\subsection{System Model}
\label{ssec:IIA}

As illustrated in Fig.~\ref{fig1}, we consider a THz downlink UM-MIMO system where a base station (BS) with widely-spaced subarrays serves a user equipment (UE) in the radiative NF region. The enlarged inter-subarray spacing not only improves the spatial multiplexing capability, but more critically, amplifies the geometric parallax effect that the same UE position induces distinct propagation distances and angles across subarrays. Specifically, the BS employs $N_t$ transmit antennas designed as $K$ uniform planar arrays (UPAs) on the x-z plane. Each subarray contains $N_s = N_t/K$ antennas with inter-element spacing $d = \lambda/2$ at central carrier wavelength $\lambda$, and the inter-subarray spacing is much larger than $d$. The UE is equipped with a UPA of $N_r$ antennas with spacing $d$.

To reduce the hardware overhead, both the BS and UE adopt HBF structure. 
The BS uses a sub-connected architecture with $L_t=K$ RF chains~\cite{yan2021joint}, therefore, the analog precoder
$\mathbf{F}_{\mathrm{RF}} \in \mathbb{C}^{N_t \times L_t}$ takes a block-diagonal form as
\begin{equation}
    \mathbf{F}_{\mathrm{RF}} = \mathrm{blkdiag}\!\left(\mathbf{f}_{\mathrm{RF},1}, \mathbf{f}_{\mathrm{RF},2}, \dots, \mathbf{f}_{\mathrm{RF},K}\right),
\end{equation}
where $\mathbf{f}_{\mathrm{RF},k} \in \mathbb{C}^{N_s}$ denotes the analog precoding vector of the $k^{\mathrm{th}}$ subarray for $k=1,\dots,K$. Since the analog precoder is implemented by phase shifters, each element in $\mathbf{f}_{\mathrm{RF},k}$ satisfies the constant-modulus constraint $|\mathbf{f}_{\mathrm{RF},k}(i)| = 1/\sqrt{N_s}$. At the UE, a fully-connected architecture is adopted, where each of the $L_r$ RF chains is connected to all $N_r$ antennas through phase shifters. The analog combiner $\mathbf{W}_{\mathrm{RF}} \in \mathbb{C}^{N_r \times L_r}$ also obeys the constant-modulus constraint.

The system operates with orthogonal frequency division multiplexing (OFDM) using $M$ subcarriers with spacing $\Delta f$. At the $m^{\mathrm{th}}$ subcarrier, the BS transmits an $N_{st}$-dimensional symbol vector $\mathbf{s}[m] \in \mathbb{C}^{N_{st}}$. It is first processed by a digital baseband precoder $\mathbf{F}_{\mathrm{BB}}[m] \in \mathbb{C}^{L_t \times N_{st}}$ and then by the analog precoder $\mathbf{F}_{\mathrm{RF}}$. After that, it is propagated through the wideband THz channel $\mathbf{H}[m] \in \mathbb{C}^{N_r \times N_t}$. The UE applies the analog combiner $\mathbf{W}_{\mathrm{RF}}$ and a digital combiner $\mathbf{W}_{\mathrm{BB}}[m] \in \mathbb{C}^{L_r \times N_{sr}}$ to form the $N_{sr}$-dimensional baseband output as
\begin{equation}
\mathbf{y}[m]=\mathbf{W}_{eq}^{\mathrm H}[m]\mathbf{H}[m]\mathbf{F}_{eq}[m]\mathbf{s}[m]+\mathbf{W}_{eq}^{\mathrm H}[m]\mathbf{n}[m],
\end{equation}
where $\mathbf{F}_{eq}[m]=\mathbf{F}_{\mathrm{RF}} \mathbf{F}_{\mathrm{BB}}[m]\in \mathbb{C}^{N_t \times N_{st}}$ and $\mathbf{W}_{eq}[m]=\mathbf{W}_{\mathrm{RF}} \mathbf{W}_{\mathrm{BB}}[m]\in \mathbb{C}^{N_r \times N_{sr}}$ represent the precoding and combining matrices, respectively, with $\big\|\mathbf{F}_{eq}[m]\big\|_F^2 = 1$ and $\big\|\mathbf{W}_{eq}[m]\big\|_F^2 = 1$, and $\mathbf{n}[m] \sim \mathcal{CN}( \mathbf{0}, \sigma_n^2 \mathbf{I}_{N_r} )$ is the additive white Gaussian noise.
Moreover, the symbol vector is normalized as $\mathbb{E}\!\left[\mathbf{s}[m]\mathbf{s}^\mathrm{H}[m]\right] = \rho\,\mathbf{I}_{N_{st}}$, where $\rho$ denotes the average transmit power per subcarrier. To satisfy the total power constraint $P_t$ across all $M$ subcarriers, we have $M \rho = P_t$.

During the pilot transmission stage for beam tracking, the above model specializes to a configuration that enables parallax extraction across subarrays which share a common local oscillator and baseband. We set $N_{st}=K$ and adopt an identity digital precoder $\mathbf{F}_{\mathrm{BB}}[m] = \frac{1}{\sqrt{K}} \mathbf{I}_K$, so that each stream is mapped to one subarray, while all subarrays are excited simultaneously toward the UE.
To ensure that the UE can resolve the individual observation from each subarray, we employ FDM pilots with disjoint subcarrier subsets $\lbrace \mathcal{M}_k \rbrace_{k=1}^K$, where only the $k^{\mathrm{th}}$ stream is active for $m\in \mathcal{M}_k$. Moreover, we adopt a comb-type allocation where each $\mathcal{M}_k$ uniformly spans the entire bandwidth with effective spacing of $K\Delta f$ for the observed tones of each subarray. 
For pilot transmission, the analog beam $\mathbf{f}_{\mathrm{RF},k}$ is selected from a standard far-field DFT codebook. Under this design, the received pilots across frequency and subarrays implicitly encode the parallax information used by the subsequent transformer-based modules.

\subsection{Near-Field Channel and Signal Structure}
\label{ssec:IIB}

For the considered THz UM-MIMO system, the Rayleigh distance $D_R=\frac{2 (S_b+S_u)^2}{\lambda}$ significantly expands, where $S_b$ and $S_u$ denote the array apertures at the BS and UE, respectively.
As a result, a typical UE operates in the NF, where a global PWM fails to capture the non-negligible wavefront curvature across widely separated subarrays, whereas a full SWM is computationally heavy. To balance accuracy and complexity, we adopt the HSPM, approximating the wavefront as planar within each subarray and spherical between subarrays.
The applicability conditions of HSPM are provided in~\cite{chen2024can,11239413}.

Within each subarray, for azimuth $\theta$ and elevation $\phi$, the generic UPA steering vector in PWM can be represented as
\begin{equation}
\mathbf{a}_N(\psi_x, \psi_z)=
\left[ 1 
\ \cdots\ 
\mathrm{e}^{j \frac{2 \pi}{\lambda} \psi_n} \ \cdots\ \mathrm{e}^{j \frac{2 \pi}{\lambda} \psi_{N-1}} 
\right]^{\mathrm{T}},
\label{eq4}
\end{equation}
where $\psi_n = d_{n_x}\psi_x + d_{n_z} \psi_z$, $\psi_x = \sin\theta\cos\phi$ and $\psi_z = \sin\phi$. Here $N$ is the number of elements in the subarray, $(d_{n_x},d_{n_z})$ denotes the distances from the $n^{\mathrm{th}}$ antenna to the reference antenna on the x- and z-axis. With $N_{\ell}$ paths between the BS and the UE, the receive and transmit steering vectors of the $k^{\mathrm{th}}$ subarray for the $\ell^{\mathrm{th}}$ path are denoted by
$\mathbf{a}_{k}^{r\ell}=
\mathbf{a}_{N_r}(\psi_{kx}^{r\ell}, \psi_{kz}^{r\ell})$ and
$\mathbf{a}_{k}^{t\ell}=
\mathbf{a}_{N_s}(\psi_{kx}^{t\ell}, \psi_{kz}^{t\ell})$, respectively. 
For the LoS path in Fig.~\ref{fig1}, the distance $D_k^1(\mathbf{p})$ and direction vector $\mathbf{u}_k^1(\mathbf{p})$ from the $k^{\mathrm{th}}$ subarray (with reference position $\mathbf{s}_k=[x_k,0,z_k]^{\mathrm T}$) to the UE (at $\mathbf{p} = [x, y, z]^{\mathrm{T}} $) are expressed as
\begin{subequations}
\label{eq:5}
  \begin{equation}
    D_k^1(\mathbf{p}) = \|\mathbf{p}-\mathbf{s}_k\|,
    \label{eq:5a}
  \end{equation}
  \begin{equation}
    \mathbf{u}_k^1(\mathbf{p}) = \frac{\mathbf{p}-\mathbf{s}_k}{D_k^1(\mathbf{p})}= [u_{k,x}, u_{k,y}, u_{k,z}]^{\mathrm{T}},
    \label{eq:5b}
  \end{equation}
\end{subequations}
In addition, the azimuth $\theta_{k}^{t}$ and elevation $\phi_{k}^{t}$ in the NF are uniquely determined by the specific geometry between $\mathbf{p}$ and $\mathbf{s}_k$, following $\cos \theta_{k}^{t} = \frac{x - x_k}{D_k^1(\mathbf{p}) \cos \phi_{k}^{t}}$ and \vspace{2pt} $\sin \phi_{k}^{t} = \frac{z - z_k}{D_k^1(\mathbf{p})}$.

However, across the widely separated subarrays, the spherical wavefront curvature is non-negligible and induces distinct phase shifts corresponding to the varying distances. Combining the PWM within subarrays and the SWM across subarrays, the frequency-domain subchannel $\mathbf{H}_k[m] \in \mathbb{C}^{N_r \times N_s}$ between the $k^{\mathrm{th}}$ subarray and the UE at subcarrier $f_m$ is modeled as
\begin{equation}
\label{eq:Hkmp}
    \mathbf{H}_k[m]
    = \sum_{\ell=1}^{N_{\ell}} \alpha_k^{\ell}\,
      {e^{-j 2\pi f_m \frac{D_k^{\ell}(\mathbf{p})}{c}}} \,
      {\mathbf{a}_{k}^{r\ell}(\mathbf{p}) \big(\mathbf{a}_{k}^{t \ell}(\mathbf{p})\big)^{\mathrm{T}}},
\end{equation}
where $\alpha_k^{\ell}$ is the complex path gain and $D_k^{\ell}(\mathbf{p})$ is the path length of the $\ell^{\mathrm{th}}$ path for the $k^{\mathrm{th}}$ subarray. The aggregate wideband channel is obtained by concatenation as $\mathbf{H}[m] = \big[ \mathbf{H}_1[m], \dots, \mathbf{H}_K[m] \big] \in \mathbb{C}^{N_r \times N_t}$, and the FDM pilot scheme in Sec.~\ref{ssec:IIA} determines which subcarriers probe each $\mathbf{H}_k[m]$.

The enlarged inter-subarray spacing amplifies the geometric parallax. Taking the first subarray as the reference and defining the geo-arm vector $\mathbf{b}_{k,1} = \mathbf{s}_k - \mathbf{s}_1$, the geometry satisfies
\begin{subequations}
\label{eq:parallax}
  \begin{equation}
    D_k(\mathbf{p}) = \sqrt{D_1^2(\mathbf{p}) + \|\mathbf{b}_{k,1}\|^2 - 2 D_1(\mathbf{p}) \big( \mathbf{u}_1(\mathbf{p}) \cdot \mathbf{b}_{k,1} \big)},
    \label{eq:7a}
  \end{equation}
  \begin{equation}
    \mathbf{u}_k(\mathbf{p}) = \frac{D_1(\mathbf{p})\mathbf{u}_1(\mathbf{p}) - \mathbf{b}_{k,1}}{D_k(\mathbf{p})},
    \label{eq:7b}
  \end{equation}
\end{subequations}
which reveals that the set $\{D_k(\mathbf{p}), \mathbf{u}_k(\mathbf{p})\}_{k=1}^K$ forms a geometry-consistent fingerprint of the UE position. In particular, the inter-subarray Time Difference of Arrival (TDoA)
$\Delta\tau_{k,1} = (D_k - D_1)/c$ and the power fingerprints reflecting path-loss variations jointly encode the UE position. Comb-FDM induces a delay periodicity $T_{\rm amb}=1/(K\Delta f)$ in $\tau_k$, but $\Delta\tau_{k,1}$ is ambiguity-free since $|\Delta\tau_{k,1}|\le d_{\max}/c < T_{\rm amb}/2$.

When the UE moves, the channel parameters evolve over time $t = qT_0$, where $q\geq1$ is the frame index and $T_0$ is the frame duration designed within the channel coherence time. Under the comb-type FDM pilot design, the pilot symbol vector on the $m^{\mathrm{th}}$ subcarrier at the $q^{\mathrm{th}}$ frame is denoted as
\begin{equation}
    \mathbf{s}[m,q] =
    \begin{cases}
        s_p[m,q]\,\mathbf{e}_k, & m\in\mathcal{M}_k,\\[1mm]
        \mathbf{0}, & \text{otherwise,}
    \end{cases}
    \label{eq:fdm_pilot_symbol}
\end{equation}
where $\mathbf{e}_k$ is the $k^{\mathrm{th}}$ canonical basis vector and $s_p[m,q]$ is a known pilot symbol with $\mathbb{E}\{|s_p[m,q]|^2\}=K\rho$. Exploiting $\mathbf{F}_{\mathrm{BB}}[m] = \frac{1}{\sqrt{K}} \mathbf{I}_K$ and the block-diagonal structure of $\mathbf{F}_{\mathrm{RF}}$, the scalar observation on the $k^{\mathrm{th}}$ stream is expressed as
\begin{equation}
    y_k[m,\!q]
    \!=\!
    \mathbf{w}_{eq,k}^{\mathrm{H}}[m]
    \mathbf{H}_k[m,\!q]\mathbf{f}_{\mathrm{RF},k}
    \frac{1}{\sqrt{K}} s_p[m,\!q]
    \!+\!n_k[m,\!q],
    \label{eq:yk_exact}
\end{equation}
where $\mathbf{w}_{eq,k}[m]$ is the effective combining vector used to extract the $k^{\mathrm{th}}$ pilot stream, and $n_k[m,q]$ is the post-combining noise. For LoS-dominant THz links and narrow pilot beams, the expression of \eqref{eq:yk_exact} can be well approximated as
\begin{equation}
    y_k[m,q] \approx \beta_k[m,q]\, e^{-j \Phi_k(m,q)} + n_k[m,q],
    \label{eq:yk_appro}
\end{equation}
where $\beta_k[m,q]$ collects path loss, beamforming gain and pilot amplitude, and is slowly varying over frequency and frames. The phase admits a frequency-linear decomposition as
\begin{equation}
    \Phi_k(m,q)
    =
    2\pi f_m \,\tau_k(q)
    + \phi_{\mathrm{ce}}(q)
    - \vartheta_k(q)
    + \epsilon_k[m,q],
    \label{eq:phi_decomp}
\end{equation}
where $\tau_k(q) = D_k^1(\mathbf{p}(q))/c$ is the propagation delay, $\phi_{\mathrm{ce}}(q)$ is the common phase error (CPE) and $\vartheta_k(q) = 2\pi \sum\limits_{i=0}^{q-1}\nu_k(i)\,T_0$ is the Doppler-induced accumulated phase with the instantaneous Doppler shift $\nu_k(q) = \frac{f_c}{c}\,\mathbf{v}^{\mathrm{T}}(q)\mathbf{u}_k(\mathbf{p}(q))$. Residual mismatch is collected in $\epsilon_k[m,q]$, including weak multipath, beam-squint effects and other model misalignment to be handled in the subsequent uncertainty-aware learning framework.

Although $\Phi_k(m,q)$ is linear with $f_m$ for fixed $q$, its evolution over $q$ is nonlinear due to time-varying Doppler and motion. For robust beam tracking, we consider the inter-frame phase increment computed from $y_k^{*}[m,q]\,y_k[m,q-1]$, yielding
\begin{equation}
    \begin{split}
        \Delta\Phi_k(m,q) &= \Phi_k(m,q)-\Phi_k(m,q-1), \quad \\[-0.2em]
        &=
        2\pi f_m\,\Delta\tau_k(q)
        \;-\;
        2\pi \bar{\nu}_k(q)\,T_0 \\[-0.2em] 
        &\quad \;+\;
        \Delta\phi_{\mathrm{ce}}(q)
        \;+\;
        \Delta\epsilon_k[m,q],
    \end{split}
    \label{eq:dphi_combined}
\end{equation}
where $\Delta\tau_k(q)$, $\Delta\phi_{\mathrm{ce}}(q)$ and $\Delta\epsilon_k[m,q]$ denote the corresponding variations over the $q^{\mathrm{th}}$ frame interval, and $\bar{\nu}_k(q)$ is the average Doppler over the $q^{\mathrm{th}}$ frame.

In summary, both $\Phi_k(m,q)$ and $\Delta\Phi_k(m,q)$ are affine in $f_m$, with slopes governed by $\tau_k(q)$ or $\Delta\tau_k(q)$.
Across subarrays, the geometric parallax constraints in \eqref{eq:parallax} tightly couple these parameters through the UE position. Therefore, the collection of frequency-linear signatures
$\{\Phi_k(m,q)\}_{k,m}$ and their increments form a geometry-consistent fingerprint of the UE. Over time, the evolution of their slopes and offsets reflects the UE dynamics. This space-frequency-time structure provides the physical foundation of our work.

\subsection{Problem Formulation}
\label{ssec:IIC}

\subsubsection{Static Parallax-Aware Localization}

We first focus on a single frame and omit its frame index for brevity. Under the comb-type FDM pilot design, stacking all pilot observations across the $K$ subarrays and their assigned subcarrier subsets $\{\mathcal{M}_k\}_{k=1}^{K}$ yields $\mathbf{y}
=
\big[
\{y_k[m]\}_{m\in\mathcal{M}_k,\;k=1}^{K}
\big]^{\mathrm T}
\in\mathbb{C}^{M}$. 
Combining the HSPM channel in~\eqref{eq:Hkmp}, the parallax constraints in~\eqref{eq:parallax} and the scalar phase structure in~\eqref{eq:yk_appro}-\eqref{eq:phi_decomp}, $\mathbf{y}$ follows the nonlinear parametric model as
\begin{equation}
\mathbf{y}=\mathbf{h}(\mathbf{p},\boldsymbol{\eta})+\mathbf{n},
\qquad
\mathbf{n}\sim\mathcal{CN}(\mathbf{0},\sigma_n^2\mathbf{I}_M),
\label{eq:y_parametric}
\end{equation}
where $\boldsymbol{\eta}$ collects nuisance parameters such as complex gains, beamforming gains and CPE. The maximum-likelihood (ML) estimator serves as a natural baseline, given by
\begin{equation}
\hat{\mathbf{p}}_{\mathrm{ML}}
=
\arg\min_{\mathbf{p}\in\mathcal{P}}
\min_{\boldsymbol{\eta}}
\big\|
\mathbf{y}-\mathbf{h}(\mathbf{p},\boldsymbol{\eta})
\big\|_2^2,
\label{eq:ml_static}
\end{equation}
where $\mathcal{P}$ is the BS coverage area. Due to the hybrid spherical-planar wavefront, inter-subarray parallax coupling and wideband comb-type sampling, \eqref{eq:ml_static} is highly non-convex and high-dimensional, making direct ML search unsuitable for low-latency beam alignment.
We therefore use PAST as a learned surrogate of the intractable ML estimator. For each frame, it maps the raw observation $\mathbf{y}$ to a compact evidence packet as
\begin{equation}
\mathcal{E}=f_{\boldsymbol{\Theta}_{\mathrm S}}(\mathbf{y}),
\end{equation}
which comprises a position estimate $\hat{\mathbf{p}}$, and parallax-aware evidence tokens extracted from the frequency-linear signatures, together with reliability indicators to quantify the frame-wise measurement quality. By design, $\mathcal{E}$ preserves the geometry-informative content of~\eqref{eq:y_parametric} while reducing the per-frame input dimension for temporal tracking.

\subsubsection{Dynamic Beam Tracking}
To capture mobility, we define the kinematic state $\mathbf{x}(q)=
\begin{bmatrix}
\mathbf{p}(q)~~
\mathbf{v}(q)
\end{bmatrix}^{\mathrm T}
\in\mathbb{R}^{6}$ and adopt a constant-velocity evolution model, given by
\begin{equation}
\mathbf{x}(q{+}1)=\mathbf{F}\mathbf{x}(q)+\mathbf{w}(q),
\qquad
\mathbf{w}(q)\sim\mathcal{N}(\mathbf{0},\mathbf{Q}),
\label{eq:state_evolution}
\end{equation}
where $\mathbf{F}$ is determined by $T_0$ and $\mathbf{Q}$ captures unmodeled accelerations. At the $q^{\mathrm{th}}$ frame, stacking the FDM pilot observations yields the observation model 
$\mathbf{y}(q)=\mathbf{h}_q(\mathbf{x}(q))+\mathbf{n}(q)$
with $\mathbf{n}(q)\sim\mathcal{CN}(\mathbf{0},\sigma_n^2\mathbf{I}_M)$.
Here, $\mathbf{h}_q(\cdot)$ encapsulates the delay, Doppler shift, the comb-type FDM sampling pattern and slowly varying gains. Equivalently, through~\eqref{eq:phi_decomp} and \eqref{eq:dphi_combined}, each subarray provides a frequency-affine phase signature whose slope and intercept evolve with $\mathbf{x}(q)$ under the geometric parallax, while residual mismatch is modeled as a disturbance.

The beam tracking task is to recursively infer $\mathbf{x}(q)$ from $\mathbf{y}(0{:}q)$. The Bayes-optimal solution would propagate the filtering posterior $p(\mathbf{x}(q)\mid\mathbf{y}(0{:}q))$ under~\eqref{eq:state_evolution} and the observation model, but exact nonlinear filtering is analytically intractable and can be computationally expensive for high-dimensional observations.
Therefore, to obtain a low-latency and robust learned tracker, we adopt a cascaded formulation as
\begin{equation}
\mathcal{E}(q)=f_{\boldsymbol{\Theta}_{\mathrm S}}(\mathbf{y}(q)),
\qquad
\hat{\mathbf{x}}(q)=f_{\boldsymbol{\Theta}_{\mathrm T}}\big(\mathcal{E}(0),\ldots,\mathcal{E}(q)\big),
\label{eq:cascade}
\end{equation}
where the TT $f_{\boldsymbol{\Theta}_{\mathrm T}}(\cdot)$ acts as a learned nonlinear filter-predictor operating on the compact evidence sequence.

This design mirrors the classical separation between measurement processing and state estimation, leveraging physics-structured tokenization and spatial-temporal attention rather than black-box modeling. Importantly, using $\mathcal{E}(q)$ avoids a monolithic spatial-temporal network over raw pilots, improving sample efficiency and reducing runtime while preserving geometry-informative signatures for tracking.

\section{PAST for Static Localization}
\label{sec:III}

In this section, we introduce the PAST, which leverages a deep neural network to extract and interpret the rich NF parallax signatures from lightweight FF codebook excitations, bypassing the need for prohibitive NF beam sweeping.

\subsection{Feature Distillation and Physical Tokenization}
\label{ssec:IIIA}

PAST takes the received comb-type FDM pilots as input and serves as a physics-informed measurement encoder. It distills geometry-informative evidence consistent with the channel model, parallax constraints, and the per-frame frequency-affine phase structure. Specifically, it compresses the high-dimensional FDM measurements into reliability-aware physical tokens, which are fused into a compact evidence packet $\mathcal{E}$ for the downstream TT. 
The pipeline below is defined for a single frame with its index omitted.
We sort $\mathcal{M}_k=\{m_{k,1},\ldots,m_{k,L_k}\}$ and define $y_{k,i}= y_k[m_{k,i}]$ and $f_{k,i}= f_{m_{k,i}}$ for $i=1,\ldots,L_k$.

To obtain stable phase-increment signatures with bounded token length and prevent unreliable patches from contaminating the wideband evidence, we partition $\{1,\ldots,L_k\}$ into $G$ disjoint contiguous groups $\{\mathcal{I}_{k,g}\}_{g=1}^{G}$. The group energy is $E_{k,g}= \sum_{i\in\mathcal{I}_{k,g}} |y_{k,i}|^2$, and the normalized power weight for $i\in\mathcal{I}_{k,g}$ is 
$w_{k,g}[i]=\frac{|y_{k,i}|^2+\varepsilon}{\sum_{j\in\mathcal{I}_{k,g}}\big(|y_{k,j}|^2+\varepsilon\big)}$,
where $\varepsilon>0$ is a small constant.
After that, the weighted frequency centroid $\bar f_{k,g}$ and spread $\sigma^2_{k,g}$ are represented as
\begin{subequations}
\label{eq:kg_combined}
  \begin{equation}
    \bar f_{k,g} = \sum_{i\in\mathcal{I}_{k,g}} w_{k,g}[i]\; f_{k,i},
    \label{eq:fbar_kg}
  \end{equation}
  \vspace{-5pt}
  \begin{equation}
    \sigma^2_{k,g} = \sum_{i\in\mathcal{I}_{k,g}} w_{k,g}[i]\; (f_{k,i}-\bar f_{k,g})^2.
    \label{eq:sigf_kg}
  \end{equation}
\end{subequations}
Since the Fisher information of $\tau_k$ scales with the power-weighted second central frequency moment, $(E_{k,g},\sigma^2_{k,g})$ are physics-grounded descriptors of groupwise delay information.

To exploit the frequency-affine structure without fragile phase unwrapping, we form adjacent-tone products. Defining
$\mathcal{I}^{\Delta}_{k,g}=\{\,i\in\mathcal{I}_{k,g}: i{+}1\in\mathcal{I}_{k,g}\}$ and
$r_{k,i}= y_{k,i+1}y_{k,i}^{*}$ for $i\in\mathcal{I}^{\Delta}_{k,g}$, under \eqref{eq:yk_appro}-\eqref{eq:phi_decomp},
we can derive
\begin{equation}
-\angle r_{k,i}
=
\Phi_{k}(f_{k,i+1})-\Phi_{k}(f_{k,i})
+\Delta\epsilon_k,
\end{equation}
where the dominant term is the delay-induced increment $2\pi(f_{k,i+1}-f_{k,i})\tau_k$, while frequency-flat phase offsets within a frame are suppressed by differencing. The remaining perturbations are absorbed into $\Delta\epsilon_k$ and reflected in the reliability score. After that, with $u_{k,i}= {r_{k,i}}/{|r_{k,i}|}$, the magnitude-weighted circular mean is expressed as
\begin{equation}
\bar u_{k,g}=
\frac{\sum_{i\in\mathcal{I}^{\Delta}_{k,g}} |r_{k,i}|\,u_{k,i}}
     {\sum_{i\in\mathcal{I}^{\Delta}_{k,g}} |r_{k,i}|}.
\label{eq:ubar}
\end{equation}
The reliability score and wrapped slope are defined as $\kappa_{k,g}=|\bar u_{k,g}|\in[0,1]$ (higher $\kappa_{k,g}$ indicates a more consistent patch) and $\varphi_{k,g}= -\angle(\bar u_{k,g})$, respectively. Since $\varphi_{k,g}$ is wrapped under the comb spacing, it is embedded continuously with
\begin{equation}
\mathbf{q}_{k,g}=
\begin{bmatrix}
\cos\varphi_{k,g}\\
\sin\varphi_{k,g}
\end{bmatrix}
=
\frac{1}{|\bar u_{k,g}|+\varepsilon}
\begin{bmatrix}
\Re\{\bar u_{k,g}\}\\
-\Im\{\bar u_{k,g}\}
\end{bmatrix}.
\label{eq:qkg}
\end{equation}
Any remaining wrapping ambiguity is handled in the subsequent attention layers by parallax consistency, cross-group evidence and reliability scores, rather than explicit unwrapping.

For each subarray-group pair $(k,g)$, a compact descriptor is represented as
\begin{equation}
\mathbf{t}_{k,g}
\!=\!
\big[
\log(E_{k,g}\!+\!\varepsilon),
\log(\sigma^2_{k,g}+\varepsilon),
\mathbf{q}_{k,g}^{\mathrm T},
\kappa_{k,g}
\big]^{\mathrm T}
\!\in\!\mathbb{R}^{d_t}.
\label{eq:tkg}
\end{equation}
Then it is mapped into the transformer latent space by
\begin{equation}
\mathbf{z}_{k,g}
=
\mathbf{W}_{emb}\,\mathbf{t}_{k,g}
+
\mathbf{e}_{k}
+
\mathbf{e}_{gr}
+
\mathbf{e}_{b},
\label{eq:zkg}
\end{equation}
where $\mathbf{W}_{emb}$ is learnable, $\mathbf{e}_{k}$ and $\mathbf{e}_{gr}$ encode the subarray and group indices, and the fixed geometry encoding $\mathbf{e}_{b}$ is derived from~\eqref{eq:parallax}. The resulting per-frame token sequence with length $KG$ is denoted as
\begin{equation}
\mathbf{Z}
=
\big[
\mathbf{z}_{1,1},\ldots,\mathbf{z}_{1,G},\;
\ldots,\;
\mathbf{z}_{K,1},\ldots,\mathbf{z}_{K,G}
\big].
\label{eq:Z}
\end{equation}

This physics-driven tokenization preserves parallax evidence while suppressing nuisance variations in raw pilots, allowing the tracker to focus on filtering and prediction under the kinematic prior, rather than low level denoising and feature learning from high dimensional complex pilots.

\subsection{Physics-Driven Attention Mechanism}
\label{ssec:IIIB}

For the token sequence $\mathbf{Z}$, PAST uses a physics-driven attention encoder to fuse multi-subarray and multi-band evidence into $\mathcal{E}$ and $\hat{\mathbf{p}}$. The attention weights explicitly incorporate the geometry prior and token-wise reliability in Sec.~\ref{ssec:IIIA}.

To quantify token-wise reliability for evidence fusion, a scalar gate $c_{k,g}\in[0,1)$ for the pair $(k,g)$ is defined as
\begin{equation}
c_{k,g}
=
\kappa_{k,g}\cdot
\sigma\!\Big(
\gamma_0
+\gamma_1\log(E_{k,g}{+}\varepsilon)
+\gamma_2\log(\sigma^2_{k,g}{+}\varepsilon)
\Big),
\label{eq:ckg}
\end{equation}
where $\sigma(\cdot)$ is the sigmoid function and $\gamma_1,\gamma_2\ge 0$ are learned scaling factors enforced by $\gamma_a=\log\!\big(1+\exp(\tilde\gamma_a)\big)$ for $a\in\{1,2\}$. It preserves the physical monotonicity that higher usable energy and larger effective frequency spread increase the confidence, yielding a bounded gate for stable scaling and gradients.
Accordingly, $c_{k,g}$ serves as a learned confidence score to steer attention toward informative and phase-consistent patches. Then, denoting the $i^{\mathrm{th}}$ element in \eqref{eq:Z} as $\mathbf{z}_i$, we form query, key and value projections $\mathbf{q}_i^{(h)}=\mathbf{W}_Q^{(h)}\mathbf{z}_i$, $\mathbf{k}_i^{(h)}=\mathbf{W}_K^{(h)}\mathbf{z}_i$ and $\mathbf{v}_i^{(h)}=\mathbf{W}_V^{(h)}\mathbf{z}_i$ in each attention head $h$. The attention logit from token $i$ to token $j$ is defined as
\begin{equation}
\ell_{ij}^{(h)}
=
\frac{\big(\mathbf{q}_i^{(h)}\big)^{\mathrm T}\mathbf{k}_j^{(h)}}{\sqrt{d_h}}
\;+\;
b_{ij}^{(h)}
\;+\;
\delta\log\big(c_{k(j),g(j)}{+}\varepsilon\big),
\label{eq:logit}
\end{equation}
where $d_h$ is the per-head dimension, $\delta\in[0,1]$ balances source selection and content aggregation to avoid over-suppression, $k(\cdot)$ and $g(\cdot)$ map each token to its subarray and group indices. The observation-independent geometry bias $b_{ij}^{(h)}$ is injected as
\begin{equation}
b_{ij}^{(h)}
=
\mathrm{MLP}_{\mathrm{geo}}^{(h)}\!\left(
\big[
(\mathbf{b}_{k(i),k(j)}/\Lambda)^{\mathrm T},
\;\Delta g_{ij}
\big]^{\mathrm T}
\right),
\label{eq:geo_bias}
\end{equation}
where $\mathbf{b}_{k(i),k(j)}=\mathbf{b}_{k,k'}$ is the known geo-arm vector, $\Lambda
=
\sqrt{\frac{2}{K(K-1)}\sum_{1\le k<k'\le K}\big\|\mathbf{b}_{k,k'}\big\|^2}$ normalizes geo-arm lengths, $\Delta g_{ij}=g(i)-g(j)$, and $\mathrm{MLP}_{\mathrm{geo}}^{(h)}(\cdot)$ is a lightweight Feed-forward network (FFN) mapping the relative-geometry features to a scalar bias. The corresponding attention weight is $\alpha_{ij}^{(h)}=\mathrm{softmax}_j(\ell_{ij}^{(h)})$, and the head output is given by
\begin{equation}
\mathbf{o}_i^{(h)}
=
\sum_{j}
\alpha_{ij}^{(h)}\,
\big(c_{k(j),g(j)}\big)^{1-\delta}\,
\mathbf{v}_j^{(h)}.
\label{eq:attn_out_past}
\end{equation}
In \eqref{eq:logit} and \eqref{eq:attn_out_past}, $\delta\log(c)$ controls competition among candidate evidence sources, while $c^{1-\delta}$ controls their contribution magnitude.
Standard multi-head concatenation and an FFN then yield updated token representations.

Motivated by \eqref{eq:phi_decomp} that all frequency groups within one subarray share the same per-frame delay and the parallax consistency across subarrays, we factorize the encoder into an intra-subarray stage $\mathrm{Enc}_{\mathrm{intra}}(\cdot)$ and an inter-subarray stage $\mathrm{Enc}_{\mathrm{inter}}(\cdot)$, both using the attention in \eqref{eq:logit}-\eqref{eq:attn_out_past}. First, $\mathrm{Enc}_{\mathrm{intra}}(\cdot)$ is applied to each $\mathbf Z_k=[\mathbf z_{k,1},\ldots,\mathbf z_{k,G}]$ to obtain $\tilde{\mathbf{Z}}_k
=[\tilde{\mathbf{z}}_{k,1},\ldots,\tilde{\mathbf{z}}_{k,G}]$, where $\mathrm{Enc}_{\mathrm{intra}}(\cdot)$ shares parameters across $k$.
A reliability-aware pooling then produces a subarray summary token as
\vspace{-2pt}
\begin{equation}
\bar{\mathbf z}_k=\sum_{g=1}^{G}\pi_{k,g}\tilde{\mathbf z}_{k,g},
\label{eq:zbar_k}
\end{equation}
where $\pi_{k,g}=\mathrm{softmax}_g(\mathbf w_{\mathcal P}^{\mathrm T}\tilde{\mathbf z}_{k,g}+\log(c_{k,g}{+}\varepsilon))$ and $\mathbf w_{\mathcal P}$ is the pooling vector. For inter-subarray gating, a subarray-level confidence score $\bar c_k=\sum_{g=1}^{G}\pi_{k,g}\,c_{k,g}$ replaces $c_{k(j),g(j)}$ in \eqref{eq:logit} and \eqref{eq:attn_out_past}, and $\mathrm{Enc}_{\mathrm{inter}}(\cdot)$ is applied to $\mathbf{S}=[\mathbf{z}_{0},\bar{\mathbf{z}}_1,\ldots,\bar{\mathbf{z}}_K]$ with $\Delta g_{ij}=0$. Here, $\mathbf{z}_{0}$ is the global token used to aggregate frame-level evidence, with $\bar c_{0}=1$ and $\mathbf{b}_{0,k}=\mathbf{b}_{k,0}=\mathbf{0}$.

The final spatial estimate is obtained by a lightweight head on the global token as
\begin{equation}
\hat{\mathbf{p}} = \mathbf{W}_p\,\mathbf{h}_{0}+\mathbf{b}_p,
\label{eq:head}
\end{equation}
where $\mathbf{h}_{0}$ is the output of $\mathbf{z}_{0}$ after $\mathrm{Enc}_{\mathrm{inter}}(\cdot)$, and $\hat{r}=\sigma\big(\mathbf{w}_r^{\mathrm T}\mathbf{h}_{0}+b_r\big)\in(0,1)$ is a compact quality indicator. After that, the per-frame evidence interface is denoted as
\begin{equation}
\mathcal{E}
=
\big[
\hat{\mathbf{p}},\;
\mathbf{h}_{0},\;
\{\bar{\mathbf{z}}_k\}_{k=1}^{K},\;
\hat{r}
\big],
\label{eq:evidence}
\end{equation}
where the fine-grained gates $\{c_{k,g}\}$ and pooling weights $\{\pi_{k,g}\}$ are auxiliary for diagnostics, and the factorized fusion reduces the dominant attention cost from $\mathcal{O}((KG)^2)$ to $\mathcal{O}(KG^2+K^2)$ per frame.

\subsection{Physics-in-the-Loop Training and Localization}
\label{ssec:IIIC}
PAST is trained as a low latency amortized solver for the ML inverse problem in \eqref{eq:y_parametric}, combining supervised localization and a physics-in-the-loop consistency regularizer.
To complete the physics loop, we define a differentiable forward operator that predicts the token evidence from a candidate position $\mathbf p$, using the same ordered-comb indices $\{f_{k,i}\}$, adjacent set $\mathcal I^\Delta_{k,g}$ and magnitude weights proportional to $|r_{k,i}|$ as in Sec.~\ref{ssec:IIIA}. For the $k^{\mathrm{th}}$ subarray with delay $\tau_k(\mathbf p)\!=\!{\|\mathbf p\!-\!\mathbf s_k\|}/{c}$,
the corresponding ideal adjacent-tone phase increment is given by
\begin{equation}
\hat r_{k,i}(\mathbf p)
=\exp\!\big(-j2\pi (f_{k,i+1}-f_{k,i})\,\tau_k(\mathbf p)\big),~~i\in\mathcal I^\Delta_{k,g},
\label{eq:rhat_p}
\end{equation}
where $|\hat r_{k,i}(\mathbf p)|=1$.
Replacing $u_{k,i}$ in \eqref{eq:ubar} with $\hat r_{k,i}(\mathbf p)$ yields the predicted group-wise circular mean $\hat{\bar u}_{k,g}(\mathbf p)$. Then, applying \eqref{eq:qkg} to $\hat{\bar u}_{k,g}(\mathbf p)$ gives the predicted wrapped-slope embedding $\hat{\mathbf q}_{k,g}(\mathbf p)$, and we define the forward map as $\mathcal F_{k,g}(\mathbf p)=\hat{\mathbf q}_{k,g}(\mathbf p)$.

To improve robustness to bad frames, we map the quality indicator $\hat r$ in \eqref{eq:head} to a log-variance $s = \log \hat\sigma_p^2
=\log\!\frac{1-\hat r+\varepsilon}{\hat r+\varepsilon}$ in a heteroscedastic regression form, and $\hat\sigma_p^2=e^s$ represents an input-dependent uncertainty of the position regression. For numerical stability, we clip $s$ to a bounded interval and use the Gaussian negative log-likelihood, given by
\begin{equation}
\mathcal L_{\mathrm{sup}}
=
\exp(-s)\,\|\hat{\mathbf p}-\mathbf p\|_2^2+s.
\label{eq:hetero_nll}
\end{equation}
To enforce physical consistency, we regularize the embedding $\mathbf q_{k,g}$ toward the forward prediction $\mathcal F_{k,g}(\hat{\mathbf p})$, using $c_{k,g}$ in \eqref{eq:ckg} as a learned precision proxy. With the wrap-safe similarity $d(\mathbf a,\mathbf b)=1-\mathbf a^{\mathrm T}\mathbf b$ for normalized embeddings, we define
\begin{equation}
\mathcal L_{\mathrm{phy}}^{\mathrm{s}}
=
\sum_{k=1}^{K}\sum_{g=1}^{G}
\frac{\mathrm{sg}\!\big(c_{k,g}\big)}{\frac{1}{KG}\sum_{k',g'}\mathrm{sg}\!\big(c_{k',g'}\big)}\,
\varpi\Big(d\big(\mathbf q_{k,g},\mathcal F_{k,g}(\hat{\mathbf p})\big)\Big),
\label{eq:Lphy}
\end{equation}
where $\varpi(\cdot)$ is a smooth robust penalty like Charbonnier and $\mathrm{sg}(\cdot)$ stops gradients to avoid trivially reducing $\mathcal L_{\mathrm{phy}}^{\mathrm{s}}$ by shrinking the gates.
The overall training objective is
\begin{equation}
\mathcal L_{\mathrm{S}}
=
\mathcal L_{\mathrm{sup}}
+\lambda_{\mathrm{phy}}^{\mathrm{s}}\,\mathcal L_{\mathrm{phy}}^{\mathrm{s}},
\label{eq:Ltotal}
\end{equation}
where $\lambda_{\mathrm{phy}}^{\mathrm{s}}$ is annealed from $0$ to its target value to stabilize early training, and weight decay is applied in the optimizer.

PAST outputs the per-frame estimate $\hat{\mathbf p}$ together with a compact uncertainty indicator, forming a reliability-aware measurement interface to the downstream temporal tracker. The algorithm of PAST is summarized in Algorithm \ref{alg:PAST}.

\begin{algorithm}[t]
\caption{PAST: Per-Frame Inference and Physics-in-the-Loop Training.}
\label{alg:PAST}
\begin{algorithmic}[1]
\STATE \textbf{Input:} Pilots $\{y_k[m]\}_{m\in\mathcal{M}_k,k=1}^{K}$, subcarriers $\{f_m\}$, geo-arms $\{\mathbf b_{k,1}\}$, partition $\{\mathcal I_{k,g}\}_{g=1}^{G}$, parameters $\boldsymbol\Theta_{\mathrm S}$.
\STATE \textbf{Output:} $\mathcal{E}
=
\big[
\hat{\mathbf{p}},\;
\mathbf{h}_{0},\;
\{\bar{\mathbf{z}}_k\}_{k=1}^{K},\;
\hat{r}
\big]$.
\STATE
\STATE {\textsc{Infer-PAST}}$(\{y_k[m]\};\boldsymbol\Theta_{\mathrm S})$
\FOR{$k=1$ \textbf{to} $K$}
  \STATE Construct $\mathbf Z_k=\{\mathbf z_{k,g}\}_{g=1}^{G}$ and $\{c_{k,g}\}_{g=1}^{G}$ via \eqref{eq:kg_combined}-\eqref{eq:ckg} (including $(\mathbf q_{k,g},\kappa_{k,g})$ in \eqref{eq:qkg}).
  \STATE $\tilde{\mathbf Z}_k \gets \mathrm{Enc}_{\mathrm{intra}}(\mathbf Z_k)$ using \eqref{eq:logit}-\eqref{eq:attn_out_past}.
  \STATE Obtain view token $\bar{\mathbf z}_k$ and confidence $\bar c_k$ via \eqref{eq:zbar_k}.
\ENDFOR
\STATE Form $\mathbf S=[\mathbf z_0,\bar{\mathbf z}_1,\ldots,\bar{\mathbf z}_K]$ with $\bar c_0=1$, and compute $\tilde{\mathbf S}\gets \mathrm{Enc}_{\mathrm{inter}}(\mathbf S)$ using \eqref{eq:logit}-\eqref{eq:attn_out_past}.
\STATE Read out $\mathbf h_0$ and predict $(\hat{\mathbf p},\hat r)$ by \eqref{eq:head}; \textbf{return} $\mathcal{E}$.
\STATE
\STATE {\textsc{Train-PAST}}$(\{(\mathbf y_n,\mathbf p_n)\};\boldsymbol\Theta_{\mathrm S})$
\STATE Run {\textsc{Infer-PAST}}; compute $\mathcal L_{\mathrm{sup}}$ by \eqref{eq:hetero_nll} and $\mathcal L_{\mathrm{phy}}^{\mathrm s}$ by \eqref{eq:rhat_p}-\eqref{eq:Lphy}.
\STATE Update $\boldsymbol\Theta_{\mathrm S}$ by minimizing $\mathcal L_{\mathrm S}$ in \eqref{eq:Ltotal}.
\end{algorithmic}
\end{algorithm}

\section{Temporal Transformer for Tracking}
\label{sec:IV}

\begin{figure*}[t!]
\centering
\includegraphics[width=1\textwidth,trim=0cm 0cm 0.05cm 0cm,clip]{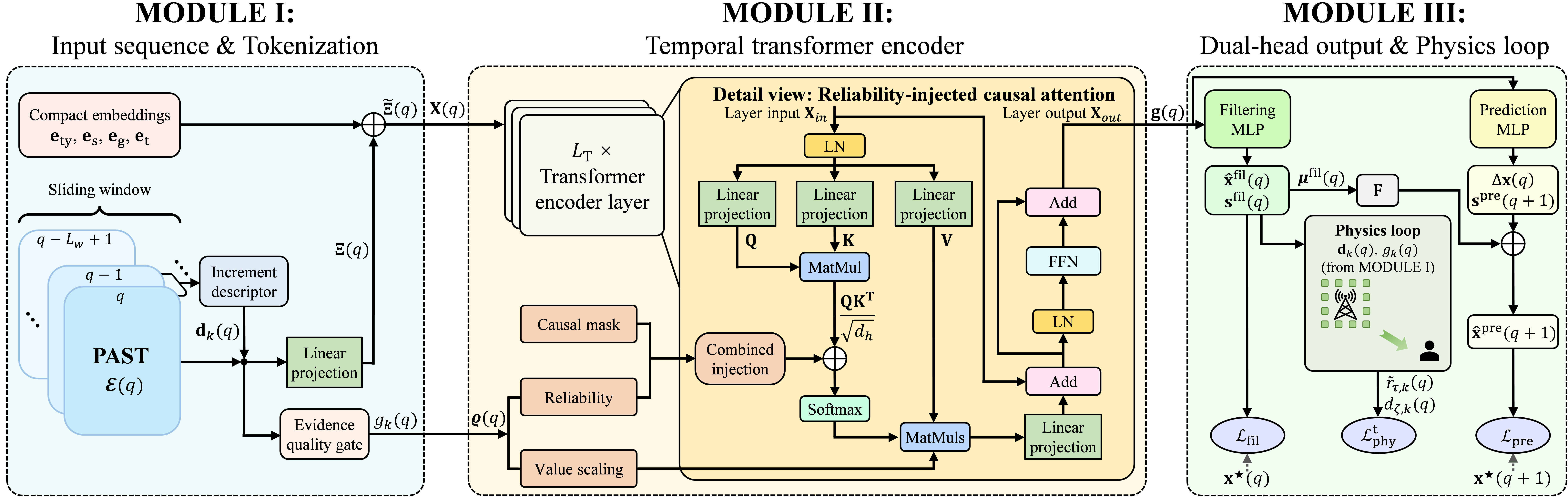}
\caption{Temporal transformer.}
\label{TT_fig} 
\end{figure*}

In this section, we introduce the causal TT that turns per-frame PAST evidence into reliability-aware tracking, jointly performing filtering and prediction under a physics loop, as shown in Fig.~\ref{TT_fig}.

\subsection{Frame-Level Design and Motion Model}
\label{ssec:IVA}

We adopt the frame interval $T_0$ defined in Sec.~\ref{ssec:IIB} and the effective pilot duration $T_{\mathrm p}\ll T_0$. Specifically, $T_{\mathrm p}$ is designed to satisfy the frame-wise quasi-static condition $2\pi f_{D,\max} T_{\mathrm p} \le \epsilon_{\mathrm p}$, where $f_{D,\max}={v_{\max}}/{\lambda}$ is the maximum Doppler shift induced by the largest UE velocity $v_{\max}$, and $\epsilon_{\mathrm p}$ bounds the intra-pilot phase rotation, ensuring the extracted evidence $\mathcal{E}(q)$ can be treated as a single state-space measurement. Moreover, the tracking update period $T_0$ is selected from a directional beam-coherence viewpoint to keep the evolution of the NF evidence between frames locally smooth. Two drift budgets, $\omega_{\max} T_0 \le \rho_\theta\,\Theta$ and $\Delta\tau_{\max} \le \rho_\tau/{B}$, are imposed, where $\Theta$ is the half-power beamwidth of the analog beam, $B$ is the bandwidth and $\rho_\theta,\rho_\tau\in(0,1)$ are dimensionless margins. The angular rate $\omega$ is upper bounded by $\omega_{\max}\le {v_{\perp,\max}}/{r_{\min}}$, where $v_{\perp,\max}$ is the maximum transverse speed and $r_{\min}$ is the minimum distance between BS and UE.
For delay drift, with $\tau_{k}(q)$ denoting the LoS propagation delay of the $k^{\mathrm{th}}$ subarray at the $q^{\mathrm{th}}$ frame, a sufficient condition is $|\tau_{k}(q)-\tau_{k}(q-1)|\le v_{r,\max}T_0/c$ for all $k$,
hence $v_{r,\max}T_0/c \le \rho_\tau/B$ ensures $\Delta\tau_{\max} \le \rho_\tau/B$.

Following Sec.~\ref{ssec:IIC}, we adopt the kinematic prior with $T_0$, and set the state transition matrix in \eqref{eq:state_evolution} as
\begin{equation}
\mathbf{F}
=
\begin{bmatrix}
\mathbf{I}_3 & T_0\mathbf{I}_3\\
\mathbf{0}   & \mathbf{I}_3
\end{bmatrix},
\label{eq:F}
\end{equation}
which induces a locally smooth propagation over one update while leaving inter-frame maneuvers to be absorbed by the downstream learnable module.
For reproducibility, the process-noise covariance is parameterized by the standard integrated acceleration form with strength $q_a\ge 0$, given by
\begin{equation}
\mathbf{Q}
=
q_a
\begin{bmatrix}
\frac{T_0^4}{4}\mathbf{I}_3 & \frac{T_0^3}{2}\mathbf{I}_3\\
\frac{T_0^3}{2}\mathbf{I}_3 & T_0^2\mathbf{I}_3
\end{bmatrix},
\label{eq:Q}
\end{equation}
where $q_a$ controls the departure from the locally smooth prior, acting as a lightweight regularizer rather than a hard motion law, and a larger $q_a$ admits stronger inter-frame deviations.
The same form applies to planar tracking by replacing $\mathbf{I}_3$ with $\mathbf{I}_2$.

\subsection{Input Sequence and Tokenization}
\label{ssec:IVB}

Following the per-frame interface in~\eqref{eq:evidence}, we index the evidence from PAST at the $q^{\mathrm{th}}$ frame as
\begin{equation}
\mathcal{E}(q)
=
\big[
\hat{\mathbf{p}}(q),\;
\mathbf{h}_0(q),\;
\{\bar{\mathbf{z}}_{k}(q)\}_{k=1}^{K},\;
\hat{r}(q)
\big],
\label{eq:Eq_def}
\end{equation}
where $\bar{\mathbf{z}}_{k}(q)$ is the view token from the $k^{\mathrm{th}}$ subarray with parallax effect and $\hat{r}(q)\in(0,1)$ is the quality indicator of each frame.
Beyond the static evidence in $\mathcal{E}(q)$, inspired by the signal structure in \eqref{eq:phi_decomp} and \eqref{eq:dphi_combined}, a slope-offset proxy via a circular least-squares fit on the unit circle is defined as
\begin{equation}
\begin{split}
&\big(\widehat{\Delta\tau}_k(q),\widehat{\zeta}_k(q)\big) = \\[-3pt]
&\arg\min_{\Delta\tau,\zeta}
\sum_{m\in\mathcal{M}_k}
w_{k,m}(q)
\Big|
e^{j\Delta\Phi_k(m,q)}
-
e^{j\big(2\pi f_m\Delta\tau-\zeta\big)}
\Big|^2,
\end{split}
\raisetag{2.3ex}
\label{eq:circ_fit}
\end{equation}
where $w_{k,m}$ is the normalized weight based on magnitude with $\sum_{m\in\mathcal{M}_k}\!\!w_{k,m}(q)\!=\!1$.
To suppress common-mode offsets shared across subarrays, we have $\widehat{\zeta}^{\mathrm{rel}}_k(q)\!=\! \widehat{\zeta}_k(q)\!-\!\widehat{\zeta}_1(q)$ for $k\!\ge\! 2$ and
$\widehat{\zeta}^{\mathrm{rel}}_1(q)\!=\! 0$.
To enable explicit detection of unreliable inter-frame increments without phase unwrapping, the residual-consistency score is defined as
\begin{equation}
\kappa^{\Delta}_k(q)
=
\Big|
\sum_{m\in\mathcal{M}_k}
w_{k,m}(q)
e^{j\big(\Delta\Phi_k(m,q)-2\pi f_m\widehat{\Delta\tau}_k(q)+\widehat{\zeta}_k(q)\big)}
\Big|.
\label{eq:kappa_delta}
\end{equation}
For each subarray, we form a compact increment descriptor as
\begin{equation}
\mathbf{d}_k(q)
=
\Big[
\frac{c}{T_0}\widehat{\Delta\tau}_k(q),\;
\cos\!\big(\widehat{\zeta}^{\mathrm{rel}}_k(q)\big),\;
\sin\!\big(\widehat{\zeta}^{\mathrm{rel}}_k(q)\big),\;
\kappa^{\Delta}_k(q)
\Big]^{\mathrm T}.
\label{eq:d_desc}
\end{equation}
For initialization, we set $\mathbf{d}_k(0)=\mathbf{0}$.

To preserve cross-subarray contrasts, we form a structured token list with $1{+}K$ tokens per frame as
\begin{equation}
\mathbf{\Xi}(q)=\big[\mathbf{\xi}_0(q),\mathbf{\xi}_1(q),\ldots,\mathbf{\xi}_K(q)\big]
\in\mathbb{R}^{(K+1)\times d_{\mathrm T}},
\label{eq:Uq_def}
\end{equation}
where the global token $\mathbf{\xi}_0(q)$ and other $\mathbf{\xi}_k(q)$ are mapped to a $d_{\mathrm T}$-dimensional space via shared linear projections, given by
\begin{subequations}
\label{eq:u_proj}
  \begin{align}
    \mathbf{\xi}_0(q)
    &=
    \mathbf{W}_0
    \big[
    \mathbf{h}_0^{\mathrm T}(q),\;
    \hat{\mathbf{p}}^{\mathrm T}(q),\;
    \hat{r}(q)
    \big]^{\mathrm T}
    +
    \mathbf{b}_0,
    \label{eq:u0_proj} 
    \\
    \mathbf{\xi}_k(q)
    &=
    \mathbf{W}_1
    \big[
    \bar{\mathbf{z}}_{k}^{\mathrm T}(q),\;
    \mathbf{d}_k^{\mathrm T}(q)
    \big]^{\mathrm T}
    +
    \mathbf{b}_1,
    \label{eq:uk_proj}
  \end{align}
\end{subequations}
where $\mathbf{W}_0,\mathbf{W}_1$ and $\mathbf{b}_0,\mathbf{b}_1$ are learnable and shared across frames.
This token grammar supports cross-view comparison and temporal linking for attention and injects physically grounded temporal drift cues via $\mathbf{d}_k(q)$ without exposing raw pilots to TT. To encode the subarray structure while keeping the temporal encoder generic, four compact embeddings are added: token type, subarray identity, geo-arm and temporal encoding.
For $k=0,\ldots,K$, we define
\begin{equation}
\tilde{\mathbf{\xi}}_k(q)
=
\mathbf{\xi}_k(q)
+
\mathbf{e}_{\mathrm{ty}}(\eta(k))
+
\mathbf{e}_{\mathrm{s}}(k)
+
\mathbf{e}_{\mathrm{g}}(k)
+
\mathbf{e}_{\mathrm{t}}(q),
\label{eq:TT_embed}
\end{equation}
where $\eta(0)=0$, $\mathbf{e}_{\mathrm{g}}(0)=\mathbf{0}$, and for $k\ge 1$, $\eta(k)=1$ with the geometry anchor $\mathbf{e}_{\mathrm{g}}(k)
=
\mathrm{MLP}_{\mathrm b}\!\left(
\mathbf{b}_{k,1}/\Lambda
\right)$ computed from the known geo-arm vectors. Here, $\mathrm{MLP}_{\mathrm b}(\cdot)$ is a two-layer perceptron with ReLU nonlinearity, output dimension $d_{\mathrm T}$ and a hidden width $\lfloor d_{\mathrm T}/2\rfloor$. In addition, $\mathbf{e}_{\mathrm{ty}}(\cdot)$ and $\mathbf{e}_{\mathrm{s}}(\cdot)$ are learnable,
and $\mathbf{e}_{\mathrm{t}}(q)$ is a standard temporal sinusoidal positional encoding. We then define a lightweight evidence-quality gate $g_k(q)=\sigma\!\big(a_0+a_r\hat{r}(q)+a_{\kappa}\kappa^{\Delta}_k(q)+\mathbf{w}_g^{\mathrm T}\mathbf{\xi}_k(q)\big)\in(0,1)$, and set $g_0(q)=1$ for the global token and learnable parameters $(a_0,a_r,a_{\kappa},\mathbf{w}_g)$.
It will be used in Sec.~\ref{ssec:IVC} to form token reliability weights and bias masked attention without altering the token construction.

For online tracking, at the $q^{\mathrm{th}}$ frame, the most recent
$L_{\mathrm w}$ evidence packets are collected to form the token tensor as
\begin{equation}
\underline{\mathbf{X}}(q)
=
\big[\tilde{\mathbf{\Xi}}(q{-}L_{\mathrm w}{+}1),\ldots,\tilde{\mathbf{\Xi}}(q)\big]
\in\mathbb{R}^{L_{\mathrm w}\times (K{+}1)\times d_{\mathrm T}},
\label{eq:window_tensor}
\end{equation}
where $\tilde{\mathbf{\Xi}}(q)\!=\![\tilde{\mathbf{\xi}}_0(q),\ldots,\tilde{\mathbf{\xi}}_K(q)]$.
To simplify the causal masking in Sec.~\ref{ssec:IVC}, \eqref{eq:window_tensor} is flattened into a token sequence with length $N=L_{\mathrm w}(K{+}1)$ in a time-major order as
\begin{equation}
\mathbf{X}(q)
=
\mathrm{Flatten}\big(\underline{\mathbf{X}}(q)\big),
\label{eq:flatten}
\end{equation}
with the index mapping $n\!\leftrightarrow \!(q'(n),k(n))$ for $n\!=\!0,\ldots,N{-}1$, where
$q'(n)=q{-}L_{\mathrm w}{+}1+\left\lfloor \frac{n}{K{+}1}\right\rfloor$, $k(n)= n \bmod (K{+}1)$, and
$\mathbf{x}_n=\tilde{\mathbf{\xi}}_{k(n)}\big(q'(n)\big)$.

With $N$ tokens per window, TT has an $\mathcal{O}(N^2)$ attention cost, independent of the subcarrier dimension due to the distilled interface in~\eqref{eq:Eq_def} and the increment descriptor in~\eqref{eq:d_desc}.

\subsection{Causal Temporal Transformer Encoder}
\label{ssec:IVC}

TT performs online causal sequence reasoning based on \eqref{eq:flatten} through masked self-attention with an additive mask on the attention logits. Accordingly, evidence from future frames is excluded. For initialization when $q<L_{\mathrm w}{-}1$, tokens with $q'(n)<0$ are padded with a padding token.
We also define the validity indicator for the $n^{\mathrm{th}}$ token as $s_n(q)=1$ for $q'(n)\ge 0$ and $s_n(q)=0$ otherwise.
Then, the additive mask matrix $\mathbf{M}(q)\in\mathbb{R}^{N\times N}$ is defined as
\begin{equation}
\mathbf{M}_{n,m}(q)
=
\begin{cases}
0, & s_m(q)\!=\!1 \ \text{and}\ s_n(q)\!=\!0,\\
0, & s_m(q)\!=\!s_n(q)\!=\!1\ \text{and}\ q'(m)\le q'(n),\\
-\infty, & \text{otherwise},
\end{cases}
\label{eq:mask_def}
\end{equation}
This construction excludes padding tokens as keys or values ($s_m\!=\!0$), prevents valid queries from attending to future frames ($q'(m)\!>\!q'(n)$) and allows bidirectional attention for tokens within each frame. The first case in \eqref{eq:mask_def} avoids an all-$-\infty$ mask row for padded queries, which would make the softmax ill-defined. These outputs are ignored at readout.

Each token $\mathbf{x}_m$ in $\mathbf{X}(q)$ is associated with a reliability weight $\varrho_m(q)\in(0,1]$, given by
\begin{equation}
\varrho_m(q)
=
\begin{cases}
1, & s_m(q)=0,\\
1, & s_m(q)=1\ \text{and}\ k(m)=0,\\
g_{k(m)}\!\big(q'(m)\big), & s_m(q)=1\ \text{and}\ 1 \leq k(m) \leq K,
\end{cases}
\label{eq:rho_def}
\end{equation}
with $\boldsymbol{\varrho}(q)=[\varrho_0(q),\ldots,\varrho_{N-1}(q)]^{\mathrm T}$.

We set $\mathbf{H}^{(0)}(q)=\mathbf{X}(q)$ and employ $L_{\mathrm T}$ identical causal encoder blocks.
For $\ell_t=1,\ldots,L_{\mathrm T}$, the $\ell_t^{\mathrm{th}}$ block follows the pre-normalization (Pre-LN) form as
\begin{subequations}
\label{eq:preln_block}
  \begin{align}
    \begin{split}
      \mathbf{H}'^{(\ell_t)}(q)
      &= \mathbf{H}^{(\ell_t-1)}(q) + \\
      &\quad \mathrm{MSA}\Big(\mathrm{LN}\big(\mathbf{H}^{(\ell_t-1)}(q)\big);\ \mathbf{M}(q),\boldsymbol{\varrho}(q)\Big),
    \end{split}
    \label{eq:preln_attn}
    \\
    \mathbf{H}^{(\ell_t)}(q)
    &= \mathbf{H}'^{(\ell_t)}(q) + \mathrm{FFN}\Big(\mathrm{LN}\big(\mathbf{H}'^{(\ell_t)}(q)\big)\Big),
    \label{eq:preln_ffn}
  \end{align}
\end{subequations}
where $\mathrm{LN}(\cdot)$ is applied row-wise, and $\mathrm{FFN}(\mathbf{z})
=
\mathrm{ReLU}\big(\mathbf{z}\mathbf{W}_1+\mathbf{b}_1\big)\mathbf{W}_2+\mathbf{b}_2$ is a two-layer position-wise FFN.
Here, $\mathbf{W}_1\in\mathbb{R}^{d_{\mathrm T}\times d_{\mathrm f}}$,
$\mathbf{W}_2\in\mathbb{R}^{d_{\mathrm f}\times d_{\mathrm T}}$, $d_{\mathrm f}=4d_{\mathrm T}$ and $\mathbf{b}_1,\mathbf{b}_2$ are learnable. With $N_h$ attention heads and $d_h=d_{\mathrm T}/N_h$, for head $h$ at layer $\ell_t$, the standard projections are expressed as
\begin{subequations}
\label{eq:qkv}
  \begin{align}
    \mathbf{Q}^{(\ell_t,h)}
    &=
    \mathrm{LN}(\mathbf{H}^{(\ell_t-1)})\mathbf{W}_Q^{(\ell_t,h)},
    \label{eq:qkv_q}
    \\
    \mathbf{K}^{(\ell_t,h)}
    &=
    \mathrm{LN}(\mathbf{H}^{(\ell_t-1)})\mathbf{W}_K^{(\ell_t,h)},
    \label{eq:qkv_k}
    \\
    \mathbf{V}^{(\ell_t,h)}
    &=
    \mathrm{LN}(\mathbf{H}^{(\ell_t-1)})\mathbf{W}_V^{(\ell_t,h)},
    \label{eq:qkv_v}
  \end{align}
\end{subequations}
where $\mathbf{W}_Q^{(\ell_t,h)},\mathbf{W}_K^{(\ell_t,h)},\mathbf{W}_V^{(\ell_t,h)}\in\mathbb{R}^{d_{\mathrm T}\times d_h}$ are learnable.
The masked attention weights are then calculated as
\begin{equation}
\begin{split}
\mathbf{A}^{(\ell_t,h)}(q)
= \mathrm{softmax}\!\Big(
&
{\mathbf{Q}^{(\ell_t,h)}\big(\mathbf{K}^{(\ell_t,h)}\big)^{\mathrm T}}\!/\!{\sqrt{d_h}}
\!+\! \mathbf{M}(q) \\
&\quad + \delta_{\mathrm T}\,\mathbf{1}\,\big[\log(\boldsymbol{\varrho}(q)+\varepsilon)\big]^{\mathrm T}
\Big),
\end{split}
\label{eq:attn_weight}
\end{equation}
where $\mathbf{1}\in\mathbb{R}^{N}$ is the all-one column vector, and $\mathrm{softmax}(\cdot)$ is applied row-wise.
The head output is computed as
\begin{equation}
\mathbf{O}^{(\ell_t,h)}(q)
=
\mathbf{A}^{(\ell_t,h)}(q)\,
\Big(\mathrm{diag}(\boldsymbol{\varrho}(q))^{1-\delta_{\mathrm T}}\mathbf{V}^{(\ell_t,h)}\Big).
\label{eq:attn_out}
\end{equation}
Therefore, TT injects $\boldsymbol{\varrho}(q)$ into attention competition and aggregation magnitude through the logit prior in \eqref{eq:attn_weight} and value scaling in \eqref{eq:attn_out}, explicitly downweighting tokens with low reliability. Finally, we obtain the multi-head output as
\begin{equation}
\mathrm{MSA}({\mathbf{H}};\mathbf{M},\boldsymbol{\varrho})
\!=\!
\big[\mathbf{O}^{(\ell_t,1)}(q),\ldots,\mathbf{O}^{(\ell_t,N_h)}(q)\big]\mathbf{W}_O^{(\ell_t)},
\label{eq:msa}
\end{equation}
where $\mathbf{W}_O^{(\ell_t)}\in\mathbb{R}^{d_{\mathrm T}\times d_{\mathrm T}}$ is learnable.
Under the time-major order in \eqref{eq:flatten}, the global token of the most recent frame has index $n_0(q)=(L_{\mathrm w}-1)(K+1)$. Its final-layer representation $\mathbf{g}(q)
=
\mathbf{H}^{(L_{\mathrm T})}_{n_0(q)}(q)\in\mathbb{R}^{d_{\mathrm T}}$ is used as the temporal summary.

\subsection{Dual-Head Output and Physics Loop}
\label{ssec:IVD}

Based on $\mathbf{g}(q)$, TT outputs a filtering estimate $\hat{\mathbf{x}}^{\mathrm{fil}}(q)$ of the current kinematic state and a one-step prediction $\hat{\mathbf{x}}^{\mathrm{pre}}(q{+}1)$ to initialize the beam tracking of the next frame. A differentiable physics loop further regularizes the outputs through NF geometry and drift-consistent increments.

The filtering head maps $\mathbf{g}(q)$ to a heteroscedastic Gaussian posterior proxy as
\begin{equation}
\big(\boldsymbol{\mu}^{\mathrm{fil}}(q),\mathbf{s}^{\mathrm{fil}}(q)\big)
=
f_{\mathrm{fil}}\!\big(\mathbf{g}(q)\big),
\label{eq:fil_head}
\end{equation}
where $\boldsymbol{\mu}^{\mathrm{fil}}(q)\!=\!
\begin{bmatrix}
\hat{\mathbf{p}}^{\mathrm{fil}}(q)~
\hat{\mathbf{v}}^{\mathrm{fil}}(q)
\end{bmatrix}^{\mathrm T}$, $\mathbf{s}^{\mathrm{fil}}(q)$ is the log-variance vector and $\mathbf{\Sigma}^{\mathrm{fil}}(q)\!=\!\mathrm{diag}(\exp(\mathbf{s}^{\mathrm{fil}}(q)))$.
For prediction, we follow the kinematic prior and learn a residual correction as
\begin{subequations}
\label{eq:pre_head}
\begin{align}
\big(\Delta\mathbf{x}(q),\mathbf{s}^{\mathrm{pre}}(q{+}1)\big)
=
f_{\mathrm{pre}}\!\big(\mathbf{g}(q)\big),
\label{eq:pre_res}
\\
\boldsymbol{\mu}^{\mathrm{pre}}(q{+}1)
=
\mathbf{F}\boldsymbol{\mu}^{\mathrm{fil}}(q)+\Delta\mathbf{x}(q),
\label{eq:pre_state}
\end{align}
\end{subequations}
where $\boldsymbol{\mu}^{\mathrm{pre}}(q{+}1)\!=\![\hat{\mathbf{p}}^{\mathrm{pre}}(q{+}1)~\hat{\mathbf{v}}^{\mathrm{pre}}(q{+}1)]^{\mathrm T}$,
$\mathbf{F}$ is given in~\eqref{eq:F} and $\mathbf{\Sigma}^{\mathrm{pre}}(q{+}1)\!=\!\mathrm{diag}(\exp(\mathbf{s}^{\mathrm{pre}}(q{+}1)))$.
With ground-truth $\mathbf{x}^{\star}(q)\!=\![\mathbf{p}^{\star}(q)~\mathbf{v}^{\star}(q)]^{\mathrm T}$, the filtering and prediction loss functions are defined as
\vspace{1pt}
\begin{subequations}
\label{eq:L_fil_pre}
\setlength{\jot}{-0.2em}
  \begin{align}
    \mathcal{L}_{\mathrm{fil}}
    &=
    \sum_{q}\!
    \Big(\!
    \big\|\mathbf{x}^{\star}(q) \! - \!\boldsymbol{\mu}^{\mathrm{fil}}(q)\big\|^2_{\big(\mathbf{\Sigma}^{\mathrm{fil}}(q)\big)^{-1}}\!
    +\!
    \log\det\mathbf{\Sigma}^{\mathrm{fil}}(q)\!
    \Big),
    \label{eq:Lfil}
    \raisetag{1.5ex}
    \\
    \begin{split}
      \mathcal{L}_{\mathrm{pre}}
      &=
      \sum_{q}\!
      \Big(\!
      \big\|\mathbf{x}^{\star}(q{+}1) \! -\!\boldsymbol{\mu}^{\mathrm{pre}}(q{+}1)\big\|^2_{\big(\mathbf{\Sigma}^{\mathrm{pre}}(q{+}1)\big)^{-1}}
      \\
      &\phantom{{}= \sum_{q}\! \Big(\!}
      +\log\det\mathbf{\Sigma}^{\mathrm{pre}}(q{+}1)
      \Big),
    \end{split}
    \label{eq:Lpre}
  \end{align}
\end{subequations}
which capture the robustness at the current frame and the forecasting capability at the next frame, respectively.

To suppress common-mode terms, we use the relative Doppler $\nu_k^{\mathrm{rel}}(\mathbf{p},\mathbf{v})\!=\!\nu_k(\mathbf{p},\mathbf{v})\!-\!\nu_1(\mathbf{p},\mathbf{v})$ for $k\!\ge\! 2$. Under the drift budgets in Sec.~\ref{ssec:IVA}, the wrap-safe proxies $\widehat{\Delta\tau}_k(q)$ and $\widehat{\zeta}^{\mathrm{rel}}_k(q)$ in~\eqref{eq:circ_fit} provide temporal cues with reliability, which are matched to the drift implied by the filtered state. We first define the normalized delay residual as
\begin{equation}
\tilde r_{\tau,k}(q)
=
\frac{\widehat{\Delta\tau}_k(q)-\big(\tau_k(\hat{\mathbf{p}}^{\mathrm{fil}}(q))-\tau_k(\hat{\mathbf{p}}^{\mathrm{fil}}(q{-}1))\big)}{\Delta\tau_{\max}},
\label{eq:rtau_norm}
\end{equation}
Then, the relative Doppler phase mismatch is modeled by
\begin{equation}
d_{\zeta,k}(q) = 1-\cos\!\big(\widehat{\zeta}^{\mathrm{rel}}_k(q)
-
2\pi T_0\,\nu_k^{\mathrm{rel}}\!\big(\hat{\mathbf{p}}^{\mathrm{fil}}(q),\hat{\mathbf{v}}^{\mathrm{fil}}(q)\big)\big),
\label{eq:dzeta}
\end{equation}
which is invariant to $2\pi$ wrapping and acts as a mild consistency regularizer. Using $g_k(q)$ as a learned reliability proxy, we form a stop-gradient weight $\omega_k(q)
=
\frac{\mathrm{sg}\scalebox{1.03}{(}g_k(q)\scalebox{1.03}{)}}{\frac{1}{K}\sum_{k'=1}^{K}\mathrm{sg}\scalebox{1.03}{(}g_{k'}(q)\scalebox{1.03}{)}}$.
In this way, the physics regularizer is obtained as
\begin{equation}
\mathcal{L}_{\mathrm{phy}}^{\mathrm{t}}
\!
=\!
\sum_{q\ge 1}\!
\left(
\sum_{k=1}^{K}\omega_k(q)\varpi\big(\tilde r_{\tau,k}^2(q)\big)
\!+\!
\sum_{k=2}^{K}\omega_k(q)\varpi\big(2d_{\zeta,k}(q)\big)\!\!
\right),
\label{eq:Lphy_TT}
\end{equation}
where $\varpi(\cdot)$ and $\mathrm{sg}(\cdot)$ are the same as in~\eqref{eq:Lphy}.
Finally, the TT training objective is expressed as
\begin{equation}
\mathcal{L}_{\mathrm{T}}
=
\mathcal{L}_{\mathrm{fil}}
+
\lambda_{\mathrm{pre}}\mathcal{L}_{\mathrm{pre}}
+
\lambda_{\mathrm{phy}}^{\mathrm{t}}\mathcal{L}_{\mathrm{phy}}^{\mathrm{t}},
\label{eq:LTT}
\end{equation}
where $\lambda_{\mathrm{phy}}^{\mathrm{t}}$ is annealed from $0$ to its target value for stable early training.

\section{Performance Evaluation}
\label{sec:V}
In this section, we evaluate the performance of the proposed PAST-TT framework for the considered THz NF scenario, and compare it with other representative methods.

\nopagebreak[4]
\subsection{Datasets and Simulation Setup}
\label{ssec:VA}

Under the system in Fig.~\ref{fig1}, we generate both the static localization dataset for PAST and the sequential tracking dataset for TT with the simulation parameters in TABLE~\ref{tab:sim_params}. In this case, $S_b\!=\!128\sqrt{10}\lambda$ and $S_u\!=\!\frac{3\sqrt{2}}{2}\lambda$, 
\pagebreak[4]\noindent yielding a Rayleigh distance $D_R\!=\!\frac{2 (S_b+S_u)^2}{\lambda}\!\approx\!331~\mathrm{m} $.
UE positions are sampled in 3D coordinates within the NF with $D_1^1(\mathbf{p}) \!\sim\! \mathcal{U}[35~\mathrm{m},120~\mathrm{m}]$. For each position, we include a dominant LoS path determined by NF geometry and two additional NLoS paths. The corresponding channel realization is generated based on QuaDRiGa~\cite{jaeckel2014quadriga}, adjusted by measurement-based statistics in urban microcell (UMi) environments obtained by our group~\cite{li2024220}, including K-factor, delay and angular spreads. In addition, the path powers are further scaled by frequency-dependent free-space loss and gaseous attenuation following ITU-R P.676. Comb-type FDM pilots are synthesized using the element-wise SWM, whereas the physics modules in Sec.~\ref{sec:II} adopt the tractable HSPM to avoid inverse crime. For tracking, trajectories are generated by the state evolution in Sec.~\ref{ssec:IVA} with frame interval $T_0$ and three mobility regimes with $q_a\!\in\!\{4,9,16\}$, where each is initialized with a random direction and speed $v\!\sim\!\mathcal U[10,30]~\mathrm{m/s}$. The above implementations ensure that the datasets preserve key NF parallax and THz propagation characteristics.

We generate $N_1=240{,}000$ static samples and $N_2=60{,}000$ trajectories, and divide them into train/validation/test sets ($80\%/10\%/10\%$) at the sample level for the static dataset and at the trajectory level for the tracking dataset.
Models are implemented in PyTorch and trained by AdamW with weight decay $10^{-2}$ and gradient clipping at $1.0$.
PAST is trained with batch size $256$ for $200$ epochs using a peak learning rate $2\times10^{-4}$ with a $5$-epoch linear warm-up followed by cosine decay to $10^{-6}$, and then it is frozen to generate $\mathcal E(q)$ for TT.
TT is trained with batch size $64$ trajectories for $150$ epochs using a peak learning rate $1\times10^{-4}$ with the same schedule, and the ground truth is not injected into its inputs. $\lambda_{\rm phy}^{\rm t}$ is linearly annealed from 0 to 0.2 over the first $30$ epochs.
Results are averaged over $R=10$ random seeds.
Experiments are conducted on a workstation equipped with $8\times$ NVIDIA GeForce RTX~4090 GPUs (CUDA 12.4; Driver 550.90.07).

\begin{table}[t]
\centering
\caption{Simulation Parameters}
\label{tab:sim_params}
\setlength{\tabcolsep}{2pt}
\renewcommand{\arraystretch}{0.95}
\begin{tabular}{c c c}
\toprule[1.5pt]
\textbf{Notation} & \textbf{Definition} & \textbf{Value} \\ 
\midrule
$f_c$ & Carrier frequency & 0.3~THz \\
$B$ & Bandwidth & 4~GHz \\
$M$ & Number of subcarriers & 1024 \\
$N_t$, $N_r$ & Number of transmit and receive antennas & 512, 16 \\
$K_x$, $K_z$ & Number of subarrays at x- and z- axis of the BS & 4, 2 \\
$L_t$, $L_r$ & Number of transmit and receive RF chains & 8, 4 \\
$d_s$ & Subarray spacing & 128$\lambda$ \\
$T_0$ & Frame interval & 1 ms \\
$q_a$ & Acceleration-noise strength in~\eqref{eq:Q} & 4, 9, 16 \\
$L_{\mathrm w}$ & Length of sliding window & 64 \\
\bottomrule[1.5pt]
\end{tabular}
\end{table}

\begin{figure}[htb]
\captionsetup{skip=3pt}
\begin{minipage}[b]{.49\linewidth}
  \centering
  \includegraphics[width=\linewidth]{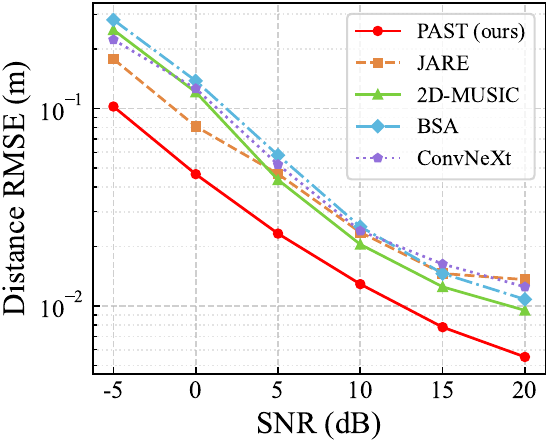}
  \vspace{1pt}
  \footnotesize (a) Distance RMSE comparison  
\end{minipage}
\hspace{0\linewidth}  
\begin{minipage}[b]{.49\linewidth}
  \centering
  \includegraphics[width=\linewidth]{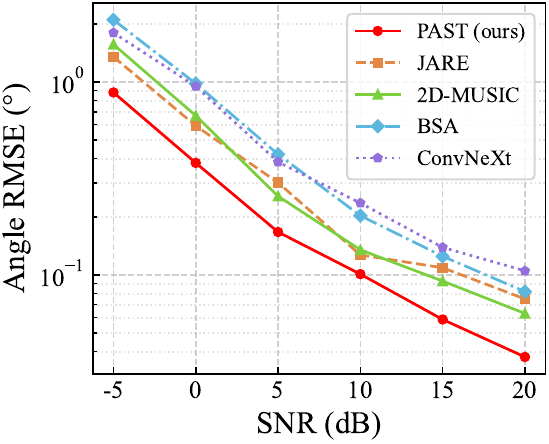}
  \vspace{1pt}
  \footnotesize (b) Angle RMSE comparison
\end{minipage}

\caption{Static RMSE comparison of different methods with SNR.}
\label{fig:static_rmse}
\end{figure}

\subsection{Performance of PAST}
\label{ssec:VB}

In this subsection, we evaluate the NF localization performance of PAST. For the static task, the UE state reduces to $\mathbf{x}=[\mathbf{p}^{\mathrm T},\mathbf{0}^{\mathrm T}]^{\mathrm T}$ and PAST outputs $\hat{\mathbf{p}}$.
Moreover, we also compare it with four representative baselines, including \emph{(i) JARE}~\cite{JARE}: a joint angle and range estimation method based on the DFT codebook, adapted by aggregating subarray-resolved pilot powers over the comb tones;
\emph{(ii) 2D-MUSIC}~\cite{2D-MUSIC}: a parametric estimator that performs a 2D spectral search on the NF manifold, with the sample covariance formed by wideband pilot snapshots;
\emph{(iii) BSA}~\cite{BSA}: a block-sparse-aware approach on a distance-dependent orthogonal dictionary, where the dominant component is recovered and mapped to the UE location through NF geometry;
\emph{(iv) ConvNeXt regression}~\cite{ConvNeXt}: a DL backbone baseline inspired by~\cite{ConvNeXt}, and it is trained to regress the distance and angles from the same FDM pilot tensor of magnitude and phase-increment features, without reproducing the CBS or TTD front-end.
All methods use the same array configuration, comb-type FDM pilots and FF DFT probing budget, and differ in the inference procedure.
The accuracy is characterized by the distance and angle RMSEs averaged over subarrays, given by
\vspace{-2pt}
\begin{subequations}
\label{eq:rmse}
  \begin{align}
    \text{RMSE}(D)
    &=\!
    \sqrt{\!
      \frac{1}{N_1 R K}\!
      \sum_{n=1}^{N_1}
      \sum_{r=1}^{R}
      \sum_{k=1}^{K}\!
      \big(
      \Delta D_{n,r}^k
      \big)^2
    },
    \label{eq:rmse_range}
    \\
    \text{RMSE}(\angle)
    &=\!
    \frac{180}{\pi}\!
    \sqrt{\!
      \frac{1}{N_1 RK}\!
      \sum_{n=1}^{N_1}
      \sum_{r=1}^{R}
      \sum_{k=1}^{K}\!
      \Big(
      \Delta \angle_{n,r}^k
      \Big)^2
    },
    \label{eq:rmse_angle}
  \end{align}
\end{subequations}
where $\Delta D_{n,r}^k=D_k(\hat{\mathbf p}_{n,r})-D_k(\mathbf p_n)$ and $\Delta \angle_{n,r}^k=\arccos\!\big(\mathbf u_k^{\mathrm T}(\mathbf p_n)\,\mathbf u_k(\hat{\mathbf p}_{n,r})\big)$.

As shown in Fig.~\ref{fig:static_rmse}, the accuracy of all methods improves with increased SNR, while PAST consistently achieves the lowest distance and angle RMSE across the tested SNR range.
At $-5$~dB, PAST attains $0.102$~m distance RMSE and $0.883^\circ$ angle RMSE, and meanwhile, the strongest baseline is JARE with RMSE of $0.178$~m and $1.35^\circ$.
At $15$~dB, the RMSE further improves to $7.81$~mm and $0.0588^\circ$, and at $20$~dB it reaches $5.52$~mm and $0.0376^\circ$.
For the high SNR regime, the best baseline is 2D-MUSIC, while PAST maintains clear gaps of $4.69$~mm and $0.0342^\circ$ at $15$~dB. 
Among the baselines, JARE is the strongest at low SNR because it estimates angle from angular support and distance from DFT probing power ratios, which is less sensitive to phase statistics, but its improvement is limited at high SNR due to a finite probing resolution. 
2D-MUSIC improves rapidly with SNR since more reliable sample covariance estimation yields cleaner subspace separation under the same pilot budget.
BSA shows a similar SNR trend, but it can be affected by mismatch in the distance-dependent dictionary and the resulting leakage in block-sparse recovery.
ConvNeXt regression also improves steadily with SNR and reaches $0.0125$~m and $0.105^\circ$ at $20$~dB, but it remains inferior to the physics-structured PAST under the same pilot budget.

\subsection{Performance of TT}
\label{ssec:VC}

This subsection evaluates TT for online filtering and prediction in beam tracking, focusing on robustness to bad frames and the resulting closed loop beam benefit.
Bad frames occur with rate $\alpha$ and are generated by applying an additional $20$~dB attenuation to the pilot observation in the corrupted frames.
We compare TT with four representative trackers, including \emph{(i) PAST+hold}, which uses the current PAST output for both filtering and prediction, \emph{(ii) EKF}, which applies an Extended Kalman Filter model with the PAST output as measurement \cite{EKF}, \emph{(iii) LSTM}, which performs causal recurrent temporal modeling for proactive tracking \cite{LSTM} under the same measurement interface, and \emph{(iv) KalmanNet}, which learns Kalman style updates under partially known dynamics \cite{kalmannet} with the same state prior. All compared methods are causal, sharing the same frozen PAST front end with the same pilot budget.

\begin{figure}[t]
\captionsetup{skip=3pt}
\begin{minipage}[b]{.49\linewidth}
  \centering
  \includegraphics[width=\linewidth]{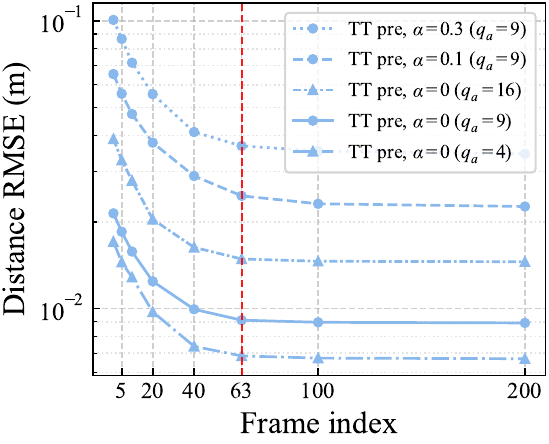}
  \vspace{1pt}
  \footnotesize (a) Distance RMSE
\end{minipage}
\hspace{0\linewidth}
\begin{minipage}[b]{.49\linewidth}
  \centering
  \includegraphics[width=\linewidth]{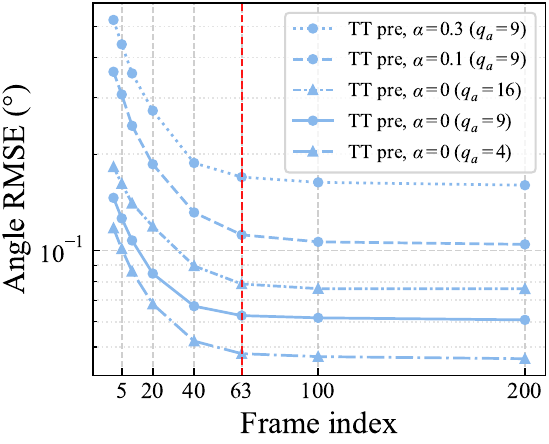}
  \vspace{1pt}
  \footnotesize (b) Angle RMSE
\end{minipage}
\caption{Time evolution of TT prediction RMSE at SNR $=15$~dB.}
\label{fig:tt_time}
\end{figure}

Fig.~\ref{fig:tt_time} shows that at SNR $={15}$~dB, TT prediction RMSE converges within a window and stays stable under bad frames and different mobilities.
At $q=200$ with $q_a=9$, the distance RMSE is $8.92$~mm for $\alpha=0$, increases to $0.0227$~m for $\alpha=0.1$, and reaches $0.0345$~m for $\alpha=0.3$.
The corresponding angle RMSE follows the same trend and converges to $0.0608^\circ$, $0.105^\circ$ and $0.160^\circ$, respectively.
With $\alpha=0$, increasing mobility from $q_a=4$ to $q_a=16$ raises the steady distance RMSE from $6.70$~mm to $0.0146$~m and the steady angle RMSE from $0.0461^\circ$ to $0.0758^\circ$.

\begin{figure}[t]
\captionsetup{skip=3pt}
\begin{minipage}[b]{.49\linewidth}
  \centering
  \includegraphics[width=\linewidth]{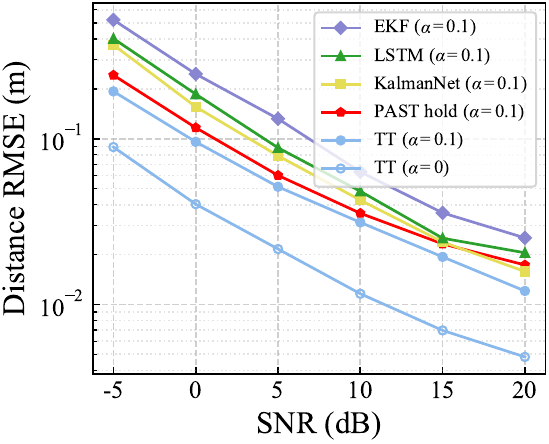}
  \vspace{1pt}
  \footnotesize (a) Distance RMSE comparison
\end{minipage}
\hspace{0\linewidth}
\begin{minipage}[b]{.49\linewidth}
  \centering
  \includegraphics[width=\linewidth]{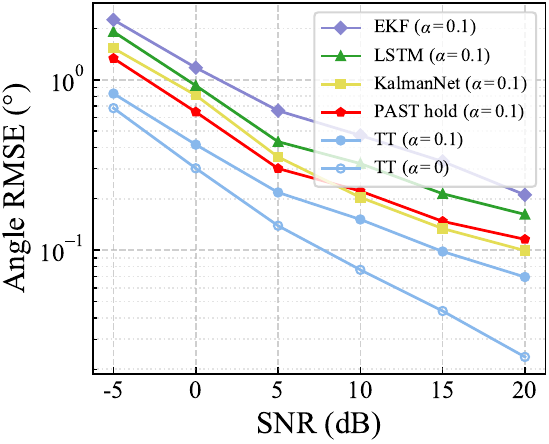}
  \vspace{1pt}
  \footnotesize (b) Angle RMSE comparison
\end{minipage}
\caption{Filtering RMSE comparison of different methods versus SNR with $\alpha=0.1$ (TT with $\alpha=0$ is shown as a reference).}
\label{fig:tt_fil_snr}
\end{figure}

Fig.~\ref{fig:tt_fil_snr} compares filtering accuracy versus SNR, where TT performs the best with its strong robustness to bad frames.
When $\alpha=0.1$ and SNR $=-5$~dB, TT attains RMSE of $0.193$~m and $0.829^\circ$, while the strongest baseline PAST+hold shows $25.4\%$ higher distance RMSE and $60.9\%$ higher angle RMSE.
EKF is the weakest with RMSE of $0.521$~m and $2.25^\circ$.
This ordering is consistent with how bad frames affect temporal inference.
PAST+hold avoids error propagation but fails to exploit temporal smoothing, while EKF is more affected once an unreliable update occurs.
KalmanNet improves upon EKF but still follows a recursive update flow \cite{kalmannet}.
At SNR $=20$~dB, TT attains $0.0121$~m and $0.0695^\circ$, with the best baseline KalmanNet exhibiting $31.2\%$ and $42.7\%$ higher distance and angle RMSE.
The reference TT curve with $\alpha=0$ reaches $4.82$~mm and $0.0236^\circ$ at SNR $=20$~dB, indicating the ceiling imposed by the bad frame process.

\begin{figure}[t]
\captionsetup{skip=3pt}
\begin{minipage}[b]{.49\linewidth}
  \centering
  \includegraphics[width=\linewidth]{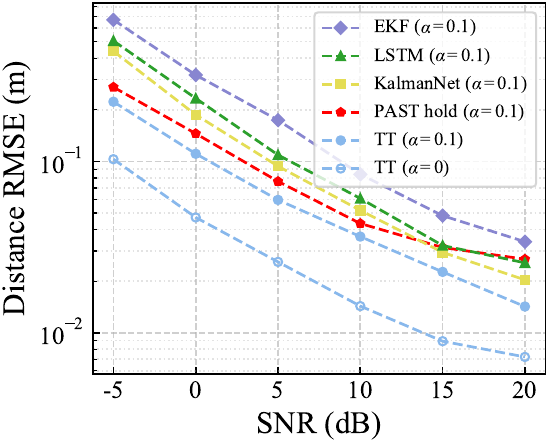}
  \vspace{1pt}
  \footnotesize (a) Distance RMSE comparison
\end{minipage}
\hspace{0\linewidth}
\begin{minipage}[b]{.49\linewidth}
  \centering
  \includegraphics[width=\linewidth]{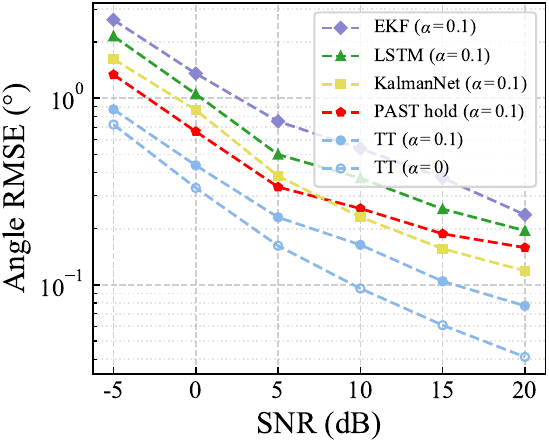}
  \vspace{1pt}
  \footnotesize (b) Angle RMSE comparison
\end{minipage}
\caption{Prediction RMSE comparison of different methods versus SNR with $\alpha=0.1$ (TT with $\alpha=0$ is shown as a reference).}
\label{fig:tt_pre_snr}
\end{figure}

Fig.~\ref{fig:tt_pre_snr} reports prediction accuracy, which determines the beam initialization for the next frame.
At SNR $=-5$~dB, TT attains $0.223$~m and $0.872^\circ$, compared with $0.271$~m and $1.34^\circ$ for PAST+hold, while EKF yields the largest RMSE of $0.670$~m and $2.63^\circ$.
At SNR $=20$~dB, TT attains $0.0142$~m and $0.0772^\circ$, while KalmanNet attains $0.0203$~m and $0.118^\circ$ as the best baseline.
Prediction is more sensitive to occasional unreliable evidence and motion changes, where TT maintains the best prediction accuracy by aggregating the causal window and downweighting unreliable evidence through its reliability-injected attention, whereas LSTM is more likely to be affected by bad frame perturbations in the input without such explicit reliability handling~\cite{LSTM}.

\begin{figure}[t]
\centering
\includegraphics[width=0.81\linewidth]{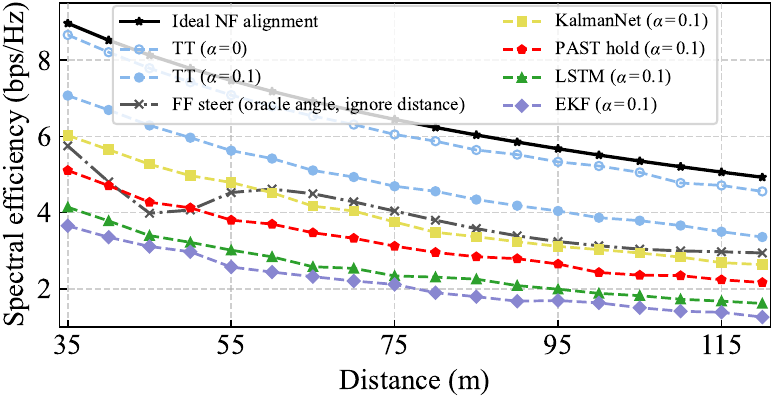}
\caption{Spectral efficiency comparison of different methods.}
\label{fig:tt_se_distance}
\end{figure}

The communication performance is evaluated by the single-stream spectral efficiency (SE), defined as 
\begin{equation}
\mathrm{SE}(q{+}1)\!=\!\frac{1}{M}\!\!\sum_{m=1}^{M}\!\log_2\!
\Big
(1\!\!+\!\!\frac{\rho\left|\mathbf w_{eq}^{\mathrm H}[m]\mathbf H[m]\mathbf f_{eq}[m]\right|^{2}}{\sigma_n^{2}}
\Big).
\label{eq:se_def}
\end{equation}
At the $q^{\mathrm{th}}$ frame, based on the prediction $\hat{\mathbf p}^{\mathrm{pre}}(q{+}1)$, an effective unit-norm transmit beam $\mathbf f_{eq}[m]$ is constructed using the HSPM in Sec.~\ref{sec:II}, with intra-subarray steering from the predicted local angles and inter-subarray phase from the predicted distances, while $\mathbf w_{eq}[m]$ is set as an ideal matched unit-norm combiner.
The ideal NF alignment curve uses the true $\mathbf p(q{+}1)$, and the FF beams discard the distance and use only the global angle. As shown in Fig.~\ref{fig:tt_se_distance},
under a fixed transmit-power budget, corresponding to $-0.1$~dBm per subcarrier, our proposed method TT stays close to the optimal SE across the tested distances, and consistently outperforms FF steering and other tracking approaches even when bad frames occur. The SE of TT with $\alpha=0$ is only $0.31$~bps/Hz and $0.37$~bps/Hz below the optimal SE at $35$~m and $120$~m, respectively. Ignoring the distance incurs up to $4.15$~bps/Hz SE loss at $45$~m which generally decreases with distance. It quantifies the importance of considering distance in NF communication.

\subsection{Ablation Study}
\label{ssec:VD}

\begin{table}[t]
\captionsetup{skip=3pt}
\caption{Ablation studies of PAST ($\alpha=0.1$, SNR$~=15~\mathrm{dB}$)}
\label{tab:ablation_past}
\centering
\footnotesize
\renewcommand{\arraystretch}{1.03} 

\begin{tabular}{c c c}
\toprule[1.5pt] 
\textbf{Method} & \textbf{Distance RMSE (m)} & \textbf{Angle RMSE ($^\circ$)} \\
\midrule 
PAST (full) & 0.0232 & 0.147 \\
w/o $c_{k,g}$ ($c_{k,g}=1$)
& 0.0351  {(+51.29\%)} & 0.239  {(+62.59\%)} \\
w/o $\mathbf{e}_b,\, b_{ij}^{(h)}$
& 0.0263  {(+13.36\%)} & 0.195  {(+32.65\%)} \\
w/o factorized fusion
& 0.0276  {(+18.97\%)} & 0.178  {(+21.09\%)} \\
w/o $\mathcal{L}_{\mathrm{phy}}^{\mathrm s}$
& 0.0329  {(+41.81\%)} & 0.203  {(+38.10\%)} \\
\bottomrule[1.5pt] 
\end{tabular}
\end{table}


\begin{table}[t]
\captionsetup{skip=3pt}
\caption{Ablation studies of TT ($\alpha=0.1$, $q_a=9$, SNR$~=15~\mathrm{dB}$)}
\label{tab:ablation_tt}
\centering
\footnotesize
\setlength{\tabcolsep}{3.5pt}
\renewcommand{\arraystretch}{0.75}  
\setlength{\aboverulesep}{2pt}
\setlength{\belowrulesep}{2pt}
\renewcommand{\makecell}[2][c]{\begin{tabular}[#1]{@{}c@{}}#2\end{tabular}}

\begin{tabular}{c c c}
\toprule[1.5pt]
\textbf{Method} 
& \makecell{\textbf{Distance RMSE (m)}\\{\footnotesize \textbf{(Filtering / Prediction)}}}
& \makecell{\textbf{Angle RMSE ($^\circ$)}\\{\footnotesize \textbf{(Filtering / Prediction)}}} \\
\midrule

TT (full)
& \makecell{0.0194\\0.0227}
& \makecell{0.0981\\0.105} \\
\midrule
w/o $\boldsymbol{\varrho}(q)$
& \makecell{0.0230  {(+18.56\%)}\\0.0293  {(+29.07\%)}}
& \makecell{0.143  {(+45.77\%)}\\0.169  {(+60.95\%)}} \\
\midrule
w/o $\mathbf{d}_k(q)$ ($a_\kappa\!=\!0$)
& \makecell{0.0226  {(+16.49\%)}\\0.0289  {(+27.31\%)}}
& \makecell{0.108  {(+10.09\%)}\\0.155  {(+47.62\%)}} \\
\midrule
w/o $\mathcal{L}_{\mathrm{phy}}^{\mathrm t}$
& \makecell{0.0228  {(+17.53\%)}\\0.0302  {(+33.04\%)}}
& \makecell{0.151  {(+53.92\%)}\\0.172  {(+63.81\%)}} \\
\midrule
w/o causal $\mathbf{M}(q)$
& \makecell{0.0210  {(+8.25\%)}\\0.0253  {(+11.45\%)}}
& \makecell{0.112  {(+14.17\%)}\\0.129  {(+22.86\%)}} \\
\midrule
w/o $f_{\mathrm{fil}}$
& \makecell{--\\0.0275  {(+21.15\%)}}
& \makecell{--\\0.163  {(+55.24\%)}} \\
\bottomrule[1.5pt]
\end{tabular}
\end{table}

Ablation studies are conducted for PAST and TT. For module-level attribution, all TT variants are trained on the same evidence sequences $\{\mathcal{E}(q)\}$ from the frozen full PAST.

In Table~\ref{tab:ablation_past}, we verify the necessity of key PAST components.
First, removing the reliability gate by setting $c_{k,g}=1$ causes the largest degradation, increasing the distance and angle RMSE by $51.29\%$ and $62.59\%$. It highlights the role of reliability-aware evidence competition and aggregation in~\eqref{eq:logit} and \eqref{eq:attn_out_past} and pooling in~\eqref{eq:zbar_k} under bad frames.
Second, without the physics-in-the-loop regularizer $\mathcal{L}_{\mathrm{phy}}^{\mathrm s}$, the RMSE increases by $41.81\%$ and $38.10\%$, demonstrating that the consistency constraint in~\eqref{eq:Lphy} provides a strong physical anchor beyond supervised regression.
Third, disabling the geometry priors $\mathbf e_b$ and $b_{ij}^{(h)}$ increases the angle RMSE by $32.65\%$, validating the geo-arm bias is essential for extracting parallax information.
Excluding the factorized fusion with a single encoder also worsens performance, suggesting that the intra- and inter-subarray factorization aligns better with the array structure while being computationally efficient.

In Table~\ref{tab:ablation_tt}, the filtering and prediction RMSEs are reported at the first and the second line of each cell, respectively. 
Without reliability-injected attention $\boldsymbol{\varrho}(q)$, the filtering and prediction angle RMSEs increase by $45.77\%$ and $60.95\%$, while removing the temporal physics loop $\mathcal{L}_{\mathrm{phy}}^{\mathrm t}$ also yields obvious degradation. These verify that strong robustness requires both reliability-aware aggregation in~\eqref{eq:attn_weight}-\eqref{eq:attn_out} and drift-consistent regularization in~\eqref{eq:Lphy_TT}. In addition, dropping the increment descriptor $\mathbf d_k(q)$ mainly affects prediction, consistent with its role of injecting inter-frame delay and offset cues from~\eqref{eq:circ_fit} to \eqref{eq:d_desc} for forecasting.
Removing the causal constraint in $\mathbf{M}(q)$ results in smaller yet consistent decline, supporting that strictly causal message passing with intra-frame bidirectional attention improves online generalization.
Disabling the filtering head $f_{\mathrm{fil}}$ degrades prediction to $0.0275$~m and $0.163^\circ$, highlighting the necessity of the dual-head design and the filtered-state anchor in~\eqref{eq:fil_head} and \eqref{eq:pre_state} for stable prediction.

\subsection{Computational Complexity}
\label{ssec:VE}

Table~\ref{tab:complexity} compares the computational complexity of different methods under the same comb-type FDM pilot budget.
For PAST, $d_{\rm P}$ is the latent width, $L_a$ and $L_b$ denote the numbers of layers in $\mathrm{Enc}_{\mathrm{intra}}(\cdot)$ and $\mathrm{Enc}_{\mathrm{inter}}(\cdot)$.
Accordingly, the dominant per-frame PAST cost is $C_{\rm PAST}=\mathcal{O}\!\left(M+L_a K G^2 d_{\rm P}+L_b K^2 d_{\rm P}\right)$. We report attention-dominated asymptotic complexity, where linear projections, FFN terms, causal masking and reliability injection only affect constant factors.
For the baselines, $P$ is the pilot-feature dimension, $N_g$ is the polar-grid size, $N_d$ is the dictionary size, $I$ is the iteration number,
$n_a$ and $n_b$ are the EKF state and measurement dimensions, $d_r$ and $d_k$ are the hidden widths of LSTM and KalmanNet, and $C_{\rm cnn}$ is the MAC count of ConvNeXt at the chosen input resolution.

\begin{table}[t]
\captionsetup{skip=3pt}
\caption{Computational Complexity Comparison}
\label{tab:complexity}
\centering
\footnotesize
\setlength{\tabcolsep}{4pt} 

\renewcommand{\arraystretch}{0.95}

\begin{tabular}{c c} 
\toprule[1.5pt]
\textbf{Method} & \textbf{Complexity} \\
\midrule
PAST (ours) & $\mathcal{O}(C_{\rm PAST})$ \\
JARE & $\mathcal{O}(P N_g)$ \\
2D-MUSIC & $\mathcal{O}(K^2 M + K^3 + N_g K^2)$ \\
BSA & $\mathcal{O}(I P N_d)$ \\
ConvNeXt regression & $\mathcal{O}(C_{\rm cnn})$ \\
\midrule
PAST+EKF & $\mathcal{O}(C_{\rm PAST}+n_a^3+n_a^2 n_b)$ \\
PAST+LSTM & $\mathcal{O}(C_{\rm PAST}+n_b d_r+d_r^2)$ \\
PAST+KalmanNet & $\mathcal{O}(C_{\rm PAST}+n_b d_k+d_k^2)$ \\
PAST+TT (ours) & $\mathcal{O}\!\left(C_{\rm PAST}+M+L_{\mathrm T}(N^2 d_{\mathrm T}+N d_{\mathrm T} d_{\mathrm f})\right)$ \\
\bottomrule[1.5pt]
\end{tabular}
\end{table}

Specifically, we use $d_{\mathrm T}=128$, $L_{\mathrm T}=4$ and $N_h=8$.
Measured on an RTX~4090, the runtime of PAST, TT frame-level tokenization and TT are $t_{\rm PAST}=0.29$~ms, $t_{\rm tok}=0.12$~ms and $t_{\rm TT}=0.32$~ms, respectively.
Scheduling PAST and frame-level tokenization in parallel, the critical-path latency is
$t_{\rm fr}=\max\{t_{\rm PAST},t_{\rm tok}\}+t_{\rm TT}=0.61$~ms.
A conservative sequential upper bound is $t_{\rm seq}=t_{\rm PAST}+t_{\rm tok}+t_{\rm TT}=0.73$~ms.
Both are below the frame interval $T_0=1$~ms in Table~\ref{tab:sim_params}, leaving sufficient margin for beam update and data transmission.

\section{Conclusion}
\label{sec:VI}

In this paper, we showed that the parallax effect can bridge low-overhead FF DFT codebook probing to accurate NF beam tracking in THz UM-MIMO without NF codebook sweeping. Comb-type FDM pilots create subarray-dependent frequency-affine phase signatures whose adjacent-tone and inter-frame increments provide delay and drift evidence while mitigating frame-wise common phase offsets. Guided by this physics-informed structure, PAST produces compact tokens with explicit reliability and delivers per-frame localization with a physics-in-the-loop consistency regularizer. TT then performs reliability-injected causal attention over a sliding window to jointly filter and predict the next-frame state for beam initialization, and it is further anchored by a temporal physics loop. Simulations confirm that at SNR 15 dB, PAST achieves 7.81 mm distance RMSE and 0.0588$^\circ$ angle RMSE, and with bad-frame rate $\alpha=0.1$, TT achieves 0.0227 m distance and 0.105$^\circ$ angle prediction RMSE with 0.61 ms critical-path latency below frame duration. Future work will address stronger NLoS and blockage, measured datasets, hardware impairments, and multi-user settings.



\vfill

\end{document}